\documentclass{aa}  
\usepackage{natbib}
\usepackage{graphicx}
\usepackage{txfonts}
\usepackage{color}
\usepackage[colorlinks=true,citecolor=blue]{hyperref}
\makeatletter
\renewcommand*\aa@pageof{, page \thepage{} of \pageref*{LastPage}}
\makeatother

\begin{document} 
  \title{Chemical evolution of elliptical galaxies with a variable IMF}
  \subtitle{A publicly available code}

  \author{Zhiqiang Yan\inst{1} \fnmsep \inst{2}
          \thanks{Emails: yan@astro.uni-bonn.de; tereza.jerabkova@eso.org; pkroupa@uni-bonn.de}
          \and
          Tereza Jerabkova \inst{1} \fnmsep \inst{2} \fnmsep \inst{3}
          \and
          Pavel Kroupa \inst{1} \fnmsep \inst{2} \fnmsep 
          \and
          Alejandro Vazdekis \inst{4}
          }

  \institute{Helmholtz-Institut f{\"u}r Strahlen- und Kernphysik (HISKP), Universität Bonn, Nussallee 14–16, 53115 Bonn, Germany
         \and
             Charles University in Prague, Faculty of Mathematics and Physics, Astronomical Institute, V Hole{\v s}ovi{\v c}k{\'a}ch 2, CZ-180 00 Praha 8, Czech Republic
        \and 
        European Southern Observatory, Karl-Schwarzschild-Straße 2, 85748 Garching bei München
        \and
Instituto de Astrofisica de Canarias, E-38200 La Laguna, Tenerife, Spain
             }

  \date{Received 6 June 2019 / Accepted 23 July 2019}

  \abstract
  {
    Growing evidence in recent years suggests a systematic variation of the stellar initial mass function (IMF), being top-heavy for starburst galaxies and possibly bottom-heavy for massive ellipticals. Galaxy chemical evolution simulations adopting an invariant canonical IMF face difficulty in simultaneously reproducing the metallicity and $\alpha$-enhancement of the massive elliptical galaxies. Applying a variable IMF that changes with time is a promising solution, however, it is non-trivial to couple a variable IMF theory with the existing galaxy evolution codes. Here we present the first open source simulation code which recalculates the galaxy-wide IMF at each time step according to the Integrated-Galactic-IMF (IGIMF) theory where the galaxy-wide IMF depends on the galactic star formation rate and metallicity. The resulting galaxy-wide IMF and metal abundance evolve with time. With this pilot work, we explore the effect of the IGIMF theory on galaxy chemical evolution in comparison with an invariant IMF.
  }


\maketitle

%________________________________________________________________

\section{Introduction}\label{sec:intro}
    
    The evolution of galaxies sensitively depends on the stellar initial mass funtion (IMF). The understanding of the IMF has been changing rapidly in the past decade. Despite a direct conflict \citep{2013pss5.book..115K} with the theoretical expectation that the IMF should vary as the star-forming environment alters (e.g., ambient gas temperature, metallicity, density, and pressure dependence as argued by \citealt{1996ApJ...464..256A,1998MNRAS.301..569L,2004MNRAS.354..367E,2007MNRAS.381L..40D,2010ApJ...720..226P}) the observed IMFs of star clusters in the local Universe are consistent with no variation \citep{2001MNRAS.322..231K,2002Sci...295...82K,2010ARA&A..48..339B,2014prpl.conf...53O,2018PASA...35...39H}. Thus the universal and invariant canonical IMF assumption was widely applied. The more recent observations, that are able to probe physical regimes further from the Solar and Galactic neighborhood, have been consistently suggesting a variation of the galaxy-wide IMF (gwIMF\footnote{See Table~\ref{tab:abbreviation} for a list of abbreviations applied in our galaxy chemical evolution model. Following \cite{2018A&A...620A..39J} we distinguish the IMF (as resulting in a star-forming event over about a Myr in a molecular cloud core) from the galaxy-wide IMF, gwIMF.}). For the distribution of low-mass stars, a bottom-heavy IMF (excess in the number of low-mass stars) in the inner regions of massive metal-rich elliptical galaxies is indicated by the galaxy mass-to-light ratio \citep{2017ApJ...838...77L}, spectral analysis of stellar-mass sensitive features \citep{1997ApJS..111..203V,2003MNRAS.340.1317V,2003MNRAS.339L..12C,2010Natur.468..940V,2012ApJ...760...71C,2013MNRAS.429L..15F,2015ApJ...806L..31M,2017MNRAS.464.3597L,2018MNRAS.477.3954P} and lensing studies \citep{2010ApJ...721L.163A,2018MNRAS.476..133O}\footnote{The bottom-heaviness of the TIgwIMF (Table~\ref{tab:abbreviation}) in the central region of massive elliptical galaxies is still under debate. See also the evidence against the bottom-heavy TIgwIMF from the latest spectral energy distribution fitting study \citep{2018MNRAS.478.4464A} and strong lensing studies \citep{2013MNRAS.434.1964S,2018MNRAS.478.1595C,2018MNRAS.481.2115S} suggesting that the current spectral analysis procedure may have unaccounted systematic uncertainties or the spectral features have a different origin.}. For the distribution of massive stars, independent evidence strongly indicates a systematically varying IMF (galaxy photometry: \citealt{2008ApJ...675..163H,2009ApJ...695..765M,2009ApJ...706..599L,2011MNRAS.415.1647G}, metal abundance of galaxy clusters: \citealt{2014MNRAS.444.3581R,2017MNRAS.470.4583U}, isotope abundance: \citealt{2017MNRAS.470..401R,2018Natur.558..260Z};), being top-heavy (more massive stars than predicted when assuming the canonical IMF) when the SFR is high and/or when the metallicity is low as summarized by \cite{2013pss5.book..115K,2017A&A...607A.126Y,2018A&A...620A..39J}. Note that the IMF can be bottom-heavy and top-heavy at the same time (see \citealt{2018A&A...620A..39J}).
    
    \begin{table*}
    \caption{The abbreviations applied in this paper.}
    \label{tab:abbreviation}
    \centering
    \begin{tabular}{ccc}
    \hline
    Abbreviation & Stands for & Note  \\ \hline
    IMF          & stellar Initial Mass Function & in embedded star clusters (star formation events in a molecular cloud) \\ 
    ECMF         & Embedded Cluster Mass Function & the initial mass distribution of the mass in stars of star clusters \\
    gwIMF        & Galaxy-Wide IMF & for a stellar population formed within a (10 Myr) star formation epoch \\
    TIgwIMF      & Time-Integrated gwIMF & the gwIMF integrated over the SFH of the galaxy \\
    SFR          & Star Formation Rate & for an entire galaxy (in the unit of M$_\odot$/yr)\tablefootmark{a} \\ 
    SFH          & Star Formation History & a function of the SFR over time \\ 
    $t_{sf}$          &  Star Formation Timescale & characteristic timescale of the primary starburst \\ 
    $f_{st}$          & Star Transformation Fraction & the mass of all stars ever formed divided by the initial gas mass\tablefootmark{b} \\ 
    \hline
    \end{tabular}
    \tablefoot{
    \tablefoottext{a}{Note that although the SFR densities differ spatially within a galaxy, the IGIMF theory has been empirically tested by the galaxy-wide-SFR--gwIMF relation as is demonstrated in \cite{2017A&A...607A.126Y}. Thus in our gwIMF calculation module (see Fig.~\ref{fig:flow_chart}), the galaxy-wide SFR is one of the input parameter.}
    \tablefoottext{b}{Different from the efficiency of star formation that links the instantaneous SFR to gas-mass (e.g., the parameter $\nu$ in \citealt{1994A&A...288...57M,1996ApJS..106..307V}; see also Section~\ref{sec:gas_flow}), $f_{st}$ is an input parameter in our code with which the initial gas mass is determined given a SFH.}
    }
    \end{table*}
    
    If the IMF is universal and invariant, the observed $\alpha$-enhancement of massive ellipticals indicate that they should have a short star formation timescale (e.g., \citealt{1994A&A...288...57M,2005ApJ...621..673T,2011MNRAS.418L..74D}). However, in the case of the most massive ellipticals, the required timescale for reproducing the $\alpha$-enhancement is too short such that the galaxy does not have enough time to recycle the metal-rich gas and enrich its stellar population \citep{2017MNRAS.466L..88D,2017MNRAS.464.4866O}. The difficulty in simultaneously reproducing the total metallicity and $\alpha$-enhancement of massive elliptical galaxies indicates the necessity of introducing at least one new degree of freedom, e.g., a varying and top-heavy gwIMF \citep{2010MNRAS.402..173A,2018MNRAS.475.3700M}.
    
    The IMF measurements in the local universe that are consistent with no variation and the apparent necessity of a variable extragalactic IMFs suggest that the IMF may vary in a complicated manner \citep{2019MNRAS.tmp..708G}. Motivated by the notion that stars form as groups in molecular cloud cores, i.e., in embedded clusters such that the gwIMF is the sum of all individual IMFs of all embedded clusters, the integrated galaxy-wide IMF (IGIMF) theory has been developed gradually by \citet{2003ApJ...598.1076K,2004MNRAS.350.1503W,2005ApJ...625..754W,2011MNRAS.412..979W,2013pss5.book..115K,2013MNRAS.436.3309W} and \citet{2018A&A...620A..39J}. The IGIMF theory suggests that the shape of the gwIMF should depend on the galactic properties at the time of the star formation and is able to explain/predict\footnote{Verified predictions include: a non-trivial relation between the mass of an embedded or a very young star cluster and the mass of the most massive star formed in this cluster \citep{2006MNRAS.365.1333W,2010MNRAS.401..275W,2013MNRAS.434...84W,2014MNRAS.441.3348W}; all the apparent isolated massive stars being back-traceable to a parent star cluster \citep{2012MNRAS.424.3037G,2017ApJ...834...94S}; integrating the observed star cluster IMFs results in the observed gwIMF \citep{2013MNRAS.436.3309W,2017A&A...607A.126Y,2018A&A...620A..39J}; a deficit of the H$\alpha$/UV signal from dwarf galaxies \citep{2009ApJ...695..765M,2009MNRAS.395..394P,2009ApJ...706..599L}.} the apparent gwIMF variation based on the molecular-cloud-core density and metallicity dependent IMF as empirically deduced by \cite{2009MNRAS.394.1529D,2012ApJ...747...72D,2012MNRAS.422.2246M} from an analysis of observed open and globular cluster and ultra-compact dwarf galaxies. Note that we distinguish in this paper between the modern and the earlier IGIMF formulation according to whether it includes the metallicity and density dependent star cluster IMF variation suggested by \cite{2012MNRAS.422.2246M} or not. A milestone is the self-consistency of the IGIMF theory considering the independent observational constraints for both the star-cluster-IMF variation and the gwIMF variation, as demonstrated in \cite{2013MNRAS.436.3309W,2017A&A...607A.126Y,2018A&A...620A..39J}.

    % a systematic error of the estimated SFR \citep{2008MNRAS.385..687W}

    When the gwIMF varies for different galaxies and at different times, galactic chemical evolution and the time-integrated gwIMF (TIgwIMF, see Table~\ref{tab:abbreviation}) will differ from the canonical estimate. Since interpretations of galaxy observations (baryonic mass, metal abundance, star formation activity, etc.) depend on the assumption of their IMF and element composition, it is important to apply the IGIMF theory to galaxy evolution simulations. However, current publicly available galaxy evolution simulation codes assume an invariant IMF (e.g., NuPyCEE, \citealt{2016ascl.soft10015R}), and it is difficult to modify the codes to allow for a varying gwIMF. A few semi-analytical models adopted the modern formulation of the IGIMF theory that include a star-cluster IMF variation (e.g., \citealt{2015MNRAS.446.3820G,2017MNRAS.464.3812F}, but without the metallicity dependence of the IMF, see Section~\ref{sec:Initial gas metallicity} below)\footnote{There are other galaxy evolution works adopting a specified star cluster IMF variation but not following \cite{2012MNRAS.422.2246M}. Such a modification of the IGIMF formulation is plausible as long as the proposed formulation is well calibrated with observation. We adopt \cite{2012MNRAS.422.2246M} as it is comprehensively calibrated but the research on star-cluster-scale IMF variation is not settled.}, however, their codes are not publicly available.
    
    This contribution makes available a public version of a galaxy chemical evolution code which allows for a constantly updated gwIMF that evolves with the galaxy metal abundance. The instantaneous gas-phase metallicity (as well as galaxy-wide SFR) determines the gwIMF of the forming stellar population, while the change of the gwIMF modifies the later metal enrichment process. To be able to explore the effect of the variable gwIMF on the chemical evolution of a galaxy, our code applies a fixed SFH and varies only the gwIMF. A full-scale hydrodynamical simulation is not necessary for this pilot study on the effect of the IGIMF theory. The aim of the presented work is to demonstrate the mechanism of how the IMF variation influences the galaxy chemical evolution and give the reader an idea of what kind of impact can be caused by such a variation. A more complicated gas-flow model without adequate observational constraints does not necessarily improve the reliability of our results.
    
    The paper is organized as follows. The IGIMF theory is outlined in Section~\ref{sec:IGIMF}. The publicly available code of our galaxy evolution model is introduced in Section~\ref{sec:code}. The assumptions applied in the model are described in Section~\ref{sec:galevo} and the systematic uncertainties of the model are discussed in Section~\ref{sec:uncertainty}.
    Example runs of the model assuming a given SFH and gwIMF formulation are shown in Section~\ref{sec:result} (and Appendix~\ref{sec:lognorm_results}) resulting in the mass composition, supernova rate, and evolution of the chemical composition. Our conclusion follows in Section~\ref{sec:conclusion}.

\section{The IGIMF theory}\label{sec:IGIMF}
    
    The fundamental insight underlying the IGIMF theory is that the systematic variation of the gwIMF, which appears to correlate with galactic SFR and metallicity, has its origin from the variation of the IMF on a molecular-cloud core, i.e., embedded star cluster scale. In other words, there exists a universal law of the star-cluster-scale IMF shape which leads to the various IMF shapes of different composite systems.
    
    Here the core ideas and assumptions of the IGIMF theory are outlined:
    
    \begin{enumerate}
    
    \item Stars always form in the dense region of the pre-star-cluster molecular cloud core with their initial masses following the IMF, thus all the stars belong to a star cluster (whether this star cluster survives or unbinds at a later time is a different issue, see \citealt{2017A&A...607A.126Y} and references therein). This simple assumption leads to a straightforward mathematical formulation and fruitful predictions which have been proved to be successful (see footnote 4).
    
    \item\label{ass:IMF} The initial mass distribution of the stars in a star cluster strictly follows an IMF\footnote{The "strictly" refers to the observed lack of Poisson scatter of the stellar mass distribution as summarized in \cite{2013pss5.book..115K}. This highly self-regulated star formation behavior implies a significant relation between the mass of stars in a star cluster and the star-cluster mass \citep{2003ApJ...598.1076K,2013pss5.book..115K,2017A&A...607A.126Y}.} where the slope of the IMF varies with the pre-star-cluster cloud core density (which is correlated with the star-cluster mass) and metallicity of the gas cloud. This IMF formulation is adopted from \citet{2012MNRAS.422.2246M}, which is currently the only comprehensive observational study that gives an empirical IMF shape dependence on both the star cluster mass and metallicity.
    
    \item\label{point:ECMF} The stellar masses of the embedded star-clusters are distributed according to the embedded star-cluster mass function (ECMF) strictly (see footnote 6). The ECMF is estimated from observations to be consistent with a single power-law function \citep[and references therein]{2003ApJ...598.1076K} with a power-law-index $\beta \approx 2$ \citep{2010MNRAS.406.1985C}\footnote{The power-law-index, $\beta$, for the ECMF is defined in \citet[their equation 9, where the Salpeter index in the same equation would be $2.35$]{2017A&A...607A.126Y}. In addition, the power-law-index is likely to depend on the galaxy-wide SFR with massive clusters forming more frequently in galaxies with a higher SFR \citep{2013MNRAS.436.3309W,2015MNRAS.446.4168R}. However, the influence of having a variable $\beta$ assumption on the galaxy chemical evolution may not be significant since it is a second-order effect.}.
    
    \item The shortest time, $\delta t$, for a galaxy to form enough star clusters that fully populate (saturate) the observed ECMF is defined as a star formation epoch. The length of the star formation epoch is likely to be a few Myr to several tens of Myr \citep[and reference therein]{2017A&A...607A.126Y}. We adopt $\delta t=10$ Myr following \citet{2004MNRAS.350.1503W,2015A&A...582A..93S}\footnote{The deduced timescale of about 10 Myr is related to the luminosity-weighted stellar age of the observed galaxies thus its SFH. The dim lower-mass stars are not well-represented in such estimations but they also contribute less to the chemical evolution and the ISM returned mass budget for having a long lifetime.}. This timescale is the key to relate the galaxy-wide SFR to the total mass of stars that populate a gwIMF (see Section~\ref{sec:time_step}).
    
    \end{enumerate}
    
    With the above axioms, it is possible to calculate the gwIMF by integrating over the IMFs of all embedded star clusters assembled in the star formation epoch, $\delta t$, in the galaxy. We note that by basing the gwIMF calculation on the empirically-constrained IMF variation, most of the implicit possible physical drivers for the variation, such as the cosmic-ray flux \citep{2010ApJ...720..226P}, are largely incorporated. Thus the IGIMF theory provides a mathematical framework that could reproduces the variety of extragalactic observations and being consistent with the local star-cluster-scale IMF measurements in so far as tests have been made \citep{2008Natur.455..641P,2013MNRAS.436.3309W,2017A&A...607A.126Y,2018A&A...620A..39J}. 
    
    We refer the readers to \citet{2017A&A...607A.126Y,2018A&A...620A..39J} for a more detailed description of the IGIMF theory, its modern mathematical formulation, as well as for a comprehensive demonstration of the resulting gwIMFs under different circumstances.
    
\section{The open source code}\label{sec:code}

\begin{figure}
        \centering
        \includegraphics[width=\hsize]{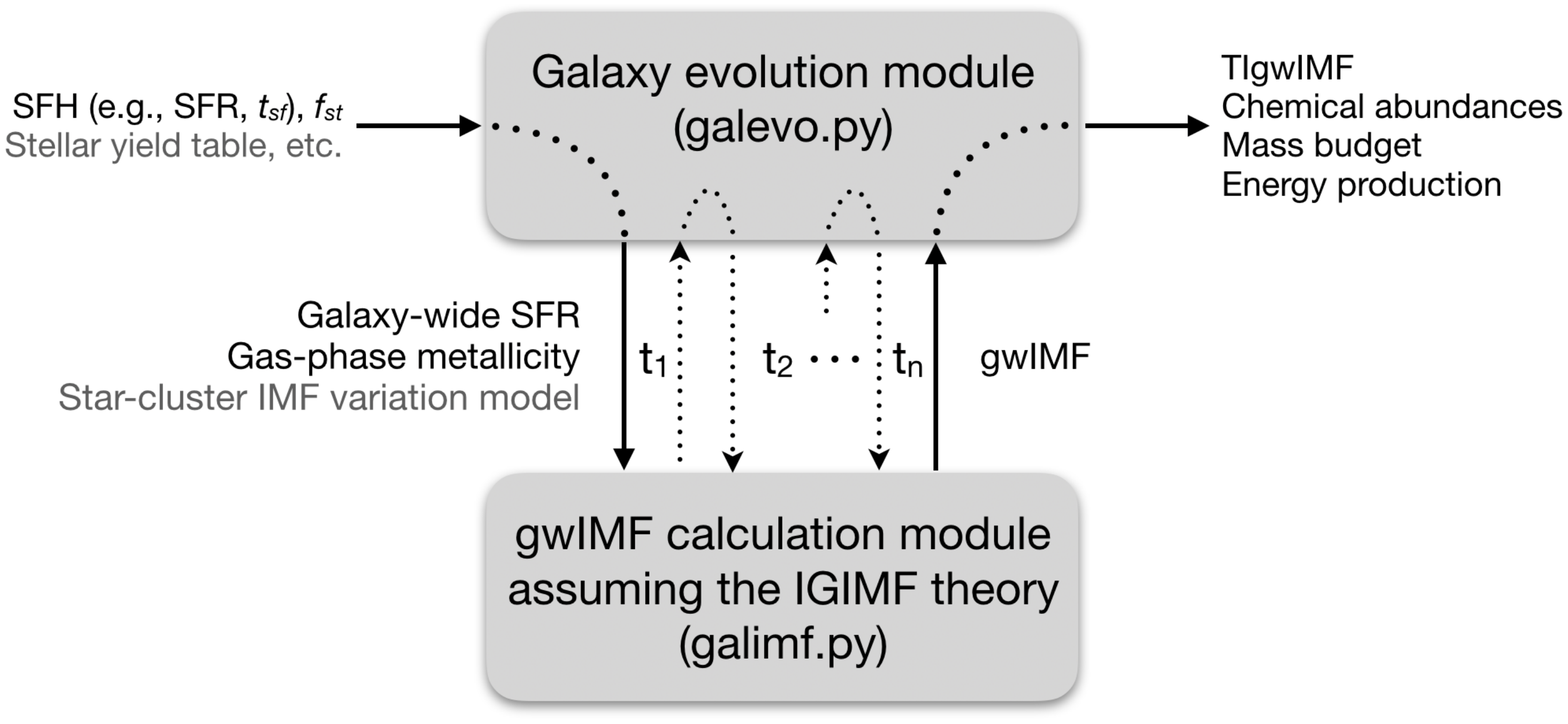}
        \caption{The interaction between the newly developed galaxy evolution module and the previously published IGIMF calculation module \citep{2017A&A...607A.126Y} which returns the gwIMF at any given time $t_i$, as well as the input/output of each module. The main inputs are shown with the black text while the default settings are shown with the gray text.}
        \label{fig:flow_chart}
    \end{figure}

    The gwIMF predicted by the IGIMF theory can be computed by the publicly available \textit{galimf.py} code in the Python3 programming language \citep{2017A&A...607A.126Y}\footnote{There is also an equivalent code in the FORTRAN programming language \citep{2019MNRAS.483...46Z} available at https://github.com/ahzonoozi/gwIMF.}. The current contribution adds to the same repository a galaxy evolution module, \textit{galevo.py}, that couples the \textit{galimf.py} code as is shown in Fig.~\ref{fig:flow_chart}. The main inputs/outputs are shown in the plot while the details can be found in the online code description.
    
    Note that the gwIMF calculation module and the Galaxy evolution module, shown as the two shadowed blocks in Fig.~\ref{fig:flow_chart}, work independently. Thus it is possible to substitute the gwIMF model assuming the IGIMF theory with another variable IMF theory or adopt the IGIMF theory into another galaxy evolution model.
    
    Both \textit{galimf.py} and \textit{galevo.py} are publicly available and maintained on GitHub  (https://github.com/Azeret/galIMF) with detailed description, instruction, and example files.
    
\section{Galaxy chemical evolution model}\label{sec:galevo}
    
    Recent research points to a two-phase galaxy formation stage, with the inner regions of present-day massive ellipticals being formed at a redshift larger than two, mostly in-situ through a quick dissipative event. This phase leads to a rather compact relic galaxy core \citep{2014ApJ...780L..20T}, which resembles the so-called red nuggets found at high redshifts \citep{2008ApJ...687L..61B,2017MNRAS.467.1929F}. This phase is followed by a significant growth in size \citep{2015ApJ...808....6H}, and to a less extent in mass, from ex-situ contributions, in a process that extends over cosmic time. These findings are recently being supported by some numerical simulations \citep{2010ApJ...725.2312O,2013MNRAS.436.3507N}. Such a scenario can be approached by a monolithic collapse that particularly applies to the central regions of giant ellipticals \citep{2006ARA&A..44..141R}.
    
    In this pilot study presenting the chemical-evolution code, we do not apply in/outflows for our default model. The influence of baryon physics bears large uncertainties and a weak predictive power \citep{2017ARA&A..55...59N,2017ApJ...835..224A}. To simulate the influence of a varying gwIMF using the IGIMF theory on galaxy evolution, we minimize the number of free parameters and therefore do not involve the hydrodynamic physics. Baryonic behavior such as inflow, cooling, mixing, selective mixing, galactic fountain, and galactic wind, are not considered (but see Section~\ref{sec:gas_flow} below for a discussion, notably that for our studied cases in- and outflows do not play significant roles) and the gas in our model is always well-mixed. All the matter is initially presented as primordial gas and the SFH is pre-specified (see Section~\ref{sec:SFH} below). 
    
    A star transformation fraction ($f_{st}$) parameter is applied being the total initial mass of stars formed over the entire galaxy formation history per unit mass of primordial gas provided. The $f_{st}$ can be adjusted to give the right metallicity of a typical galaxy and is invariant for all simulations in this paper. A higher/lower $f_{st}$ leads to a higher/lower final metallicity of the simulated galaxies. Thus a good fit on the metallicity of a single galaxy does not give a simulation much credit but the ability to fit the metallicity--galaxy-mass relation can constrain the choice of the gwIMF theory.
    
    We do not modify the metal yield table given by the stellar evolution and explosion studies\footnote{Earlier galaxy chemical evolution studies suggest a modification of the literature yield table in order to fit the observation but lack independent support \citep{2004MNRAS.347..968P,2009A&A...499..711R}.}. Different stellar yield tables are available in the code while the applied one for this paper is specified in Section~\ref{sec:Metal_enrichment} below.
    
    Here the basic components of our galaxy chemical evolution model are described:
    
    \subsection{Simulation time step}\label{sec:time_step}
    
    We aim to simulate a continuous star formation activity. The shortest age difference between different stars is set to be 10 Myr, about 1\% of the nominal galaxy formation timescale of the massive elliptical galaxies. The 10 Myr is preferred as it is the characteristic timescale of the formation of a star-cluster population applied by the IGIMF theory (see \citealt{2004MNRAS.350.1503W,2017A&A...607A.126Y} and Section~\ref{sec:IGIMF} above). Thus all stars formed within a $\delta t=10$ Myr time grid are assigned the same age.
    
    Note that we do not preform a dynamical simulation and the chemical abundance at any time is determined solely by the former stellar populations. Thus it is not necessary to do the calculation at each 10 Myr during the time period without star formation activity. Other than the time span that has star formation, only a few additional time steps are inserted manually since we want the galaxy evolution outcomes to extend to about 13 Gyr even if the star formation quenches within the first Gyr.
    
    At a new time step, $t_{i}$, without star formation activity, the gas-phase element abundances are calculated according to the values at the nearest previous time step, $t_{i-1}$, and all the ejections from dying stars and type Ia supernovae (SNIa) during time $t_{i-1}$ to $t_{i}$. If there is star formation at time $t_{i}$, the new stellar population is stamped with the current gas-phase metallicity as its initial stellar metallicity. The total mass of the stellar population is given by the galaxy-wide SFR specified for this epoch (in the unit of M$_\odot$/yr) times the 10 Myr; and the gwIMF of this star formation epoch is given by the IGIMF theory that depends on the galaxy-wide SFR and the gas-phase metallicity at $t_{i}$ (see Section~\ref{sec:IGIMF}). Thus a new stellar population is generated and the amount of mass for every element transformed into stars is deducted from the gas-phase. Such a calculation is performed for every time step that has star formation and at additional time steps, if specified, that do not have star formation.
    
    \subsection{Star Formation History}\label{sec:SFH}
    
    In our code, there are two different modes to determine the SFH, either by a relation between the SFR with the instantaneous gas-mass or using a specified SFH. The two modes require different input parameters and this paper only demonstrates the latter mode\footnote{With a top-heavy IMF, a new stellar population returns most of its mass to the gas-phase shortly after its formation. In this case, the gas mass is hardly depleted. In order to quench the star formation in the former mode, one has to introduce other considerations and assumptions, such as the energy produced by supernovae and the binding energy of a galaxy, which is out of the scope of the current contribution.}.
    
\begin{figure}
        \centering
        \includegraphics[width=\hsize]{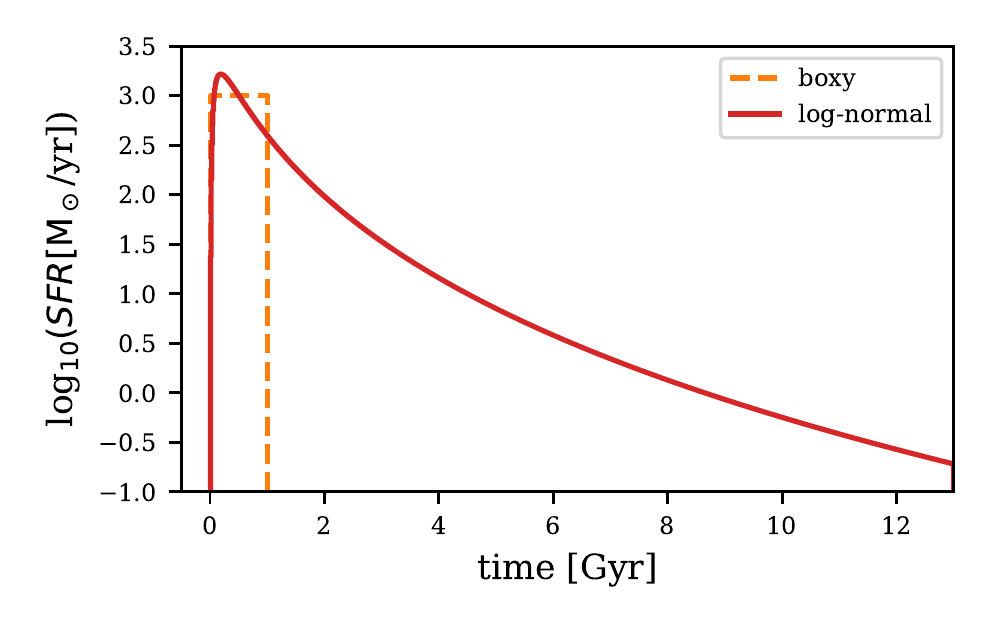}
        \caption{Adopted SFHs. The galaxy evolution simulation comparing different IMF assumptions (Section~\ref{sec:result}) adopts the boxy SFH (orange dashed line) with a constant $SFR=10^3$ M$_\odot$/yr and a $t_{sf}$ of 1 Gyr. The simulation adopting the log-normal SFH (solid red line) finishes its $50\%$, $75\%$, and $90\%$ star formation (in mass fraction) in 0.5, 0.98, and 1.8 Gyr, respectively. A comparison of the galaxy chemical evolution results adopting these two different SFHs are shown in Appendix~\ref{sec:lognorm_results}.}
        \label{fig:SFH}
    \end{figure}

    The SFH is set up before the simulation starts where the SFR for each time step, $SFR$, is specified. For example, the boxy and log-normal distributions applied for this contribution are shown in Fig.~\ref{fig:SFH}.
    
    Previous studies assuming a boxy SFH (or applying a galactic wind which assumes no star formation activity after a certain time) and taking into consideration the age and metallicity dependence on the stellar luminosity in the observations conclude that there is no notable difference between the luminosity- and mass-weighted metal abundance \citep{1998A&A...335..855M,1999MNRAS.302..537T,2009A&A...499..711R}. This claim may no longer be valid under an extended SFH with recent star formation activity. As is discussed in Section~\ref{sec:Recent_star_formation} below, the luminosity-weighted abundances can be significantly different from the mass-weighted results even if the SFR at late time is much smaller than the SFR of the initial starburst in the case of our log-normal SFH model.
    
    An important aspect of the SFH is the time-scale over which the burst or galaxy formation occurred ($t_{sf}$, see Table~\ref{tab:abbreviation}). The $t_{sf}$ is understood to be shorter for more massive elliptical galaxies with a $t_{sf}$ of about 1 Gyr for the most massive ellipticals with the strongest indication come from their $\alpha$-enhancement \citep{2004MNRAS.347..968P,2009A&A...505.1075P,2005ApJ...621..673T,2010MNRAS.404.1775T}. This relation is referred to as the downsizing of the star-formation timescale. However, note that the estimation of $t_{sf}$ according to the observed [Mg/Fe] value would be different with a different IMF assumption \citep{1994A&A...288...57M,1999MNRAS.302..537T,2009A&A...499..711R,2016MNRAS.456L.104M}. In this first contribution, we aim to discuss only the influence of the different IMF assumptions on the chemical evolution of massive elliptical galaxies and assume $t_{sf}$ to be 1 Gyr. An analysis of the possible variation of $t_{sf}$ will be covered in our next paper.
    
    \subsection{Gas-, star-, and remnant-mass evolution}
    
\begin{figure}
        \centering
        \includegraphics[width=\hsize]{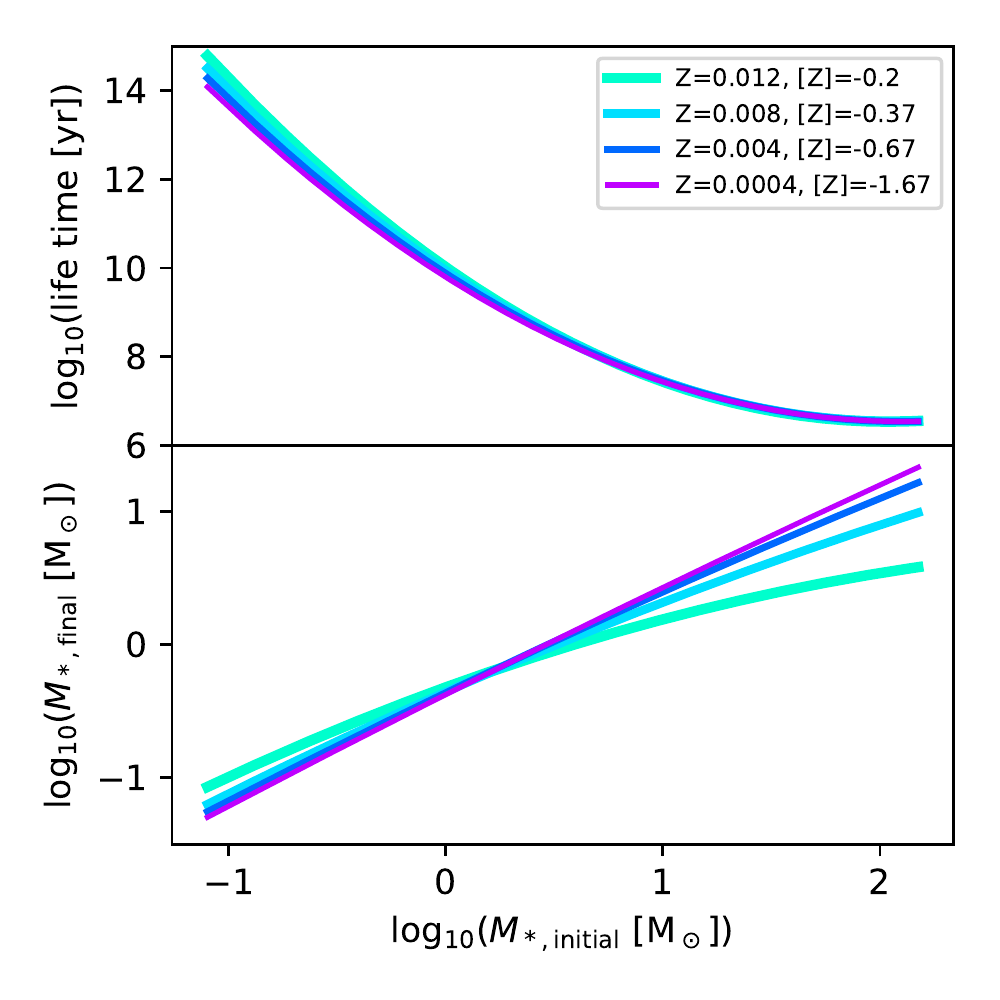}
        \caption{The lifetime (upper panel) and remnant mass, $M_{*,final}$ (lower panel), of a star as a function of its initial mass, $M_{*,\rm initial}$, and metal mass fraction, Z (or [Z] as defined in Eq.~\ref{eq:Z}). The shown relations are one-dimensional smoothing spline fits to the stellar evolution tables given by \cite{2001A&A...370..194M} and \cite{1998A&A...334..505P} for AGB and massive stars, respectively. The color code is the same as in Fig.~\ref{fig:1000_IMF_evolution}, and \ref{fig:extended_1000_IMF_evolution}.}
        \label{fig:stellar_lifetime_final_mass}
    \end{figure}

    The mass of different elements in gas, living stars, and the total stellar remnants mass are updated at each time step in our galaxy evolution model. The mass in gas transforms into stars according to the assumed SFH. The lifetime of a star and the mass of its stellar remnant is taken from the stellar evolution tables of \cite{1998A&A...334..505P} and \cite{2001A&A...370..194M} as a function of stellar initial mass and metallicity\footnote{Data pre-processing done by Christian, 25 Feb 2016, for the Public NuGrid Python Chemical Evolution Environment (NuPyCEE, https://github.com/NuGrid/NuPyCEE). The data given by the NuPyCEE program is adopted and shown in our Fig.~\ref{fig:stellar_lifetime_final_mass} and \ref{fig:steller_yields}. Note that the net yield from \cite{2001A&A...370..194M} is converted to total yield by NuGrid initial composition files and it has been updated. The newest version updated on May 2018 shows no significant difference compared with the old version we applied. We keep using the old version since it covers larger mass and metallicity ranges.}. The living stars eject gas and become stellar remnants after exhausting their lifetime. The gas reservoir and the stellar ejected gas are instantaneously well-mixed at the time of ejection.
    
    Following \citet{2018ApJS..237...42R}, a smoothing spline fit on the logarithmic scale is applied to the adopted relation of the stellar lifetime and remnant-mass versus stellar initial mass. As is shown in Fig.~\ref{fig:stellar_lifetime_final_mass}, massive stars have shorter lifetimes and metal-rich stars have smaller remnant masses. The fitted lifetime instead of the original values given by the stellar evolution model is applied in our galaxy evolution model. This treatment not only prevents the galaxy evolution model to be sensitive to the non-monotonic fluctuations of the stellar lifetime and remnant-mass given by the stellar model with large uncertainties; but also unites (and smooths the transition between) the values of intermediate-mass stars given by \cite{2001A&A...370..194M} and the values of massive stars given by \cite{1998A&A...334..505P}. We note that the spline fitted remnant mass (Fig.~\ref{fig:stellar_lifetime_final_mass}, lower panel) is only used to estimate the remnant mass (shown by the dashed "all remnants" line in Fig.~\ref{fig:mass_evolution} below) and does not modify the amount of mass ejection of the dying stars from their original values. The mass difference between the stellar initial and final mass is considered as ejected gas at the time when the star dies, while all SNIa are assumed to increase the gas-phase mass (and decrease the remnant mass) by 1.15 M$_\odot$ as is given by \cite{1993oee..conf..297T}\footnote{The value is adopted from \citet[their TNH93 dataset]{1997MNRAS.290..623G}. The 1.15 M$_\odot$ comes from summing over the Fe, Si, O, S, Mg, and Ne element ejections.}. Stellar mass loss due to the stellar wind is accounted for when the star dies. The current total mass of living stars (the solid lines in Fig.~\ref{fig:mass_evolution} below) is approximated by the sum of all the stellar initial masses of the stars that are still living.
    
    \subsection{Metal enrichment}\label{sec:Metal_enrichment}
    
    The gas-phase metallicity is set to have a primordial composition initially following the element abundances given by \cite{2016RvMP...88a5004C}\footnote{Since it is not allowed to have an element with zero mass in our logarithmic scale output, we apply an element abundance of $10^{-6}$ times the Solar value if it should be zero according to the big bang nucleosynthesis.} and then the gas phase is enriched by the ejected gas from dying stars and SNIa events. The metal-rich gas forms stars and elevates the average stellar metallicity. 
    
    Since the formation scenario of the SNIa event is still unsettled, we assume all SNIa have the same yield (as is also assumed in other studies, e.g., \citealt{2014MNRAS.445..970D}) given by \citet[their TNH93 dataset]{1997MNRAS.290..623G}. That is, the SNIa yield neither depends on the progenitor mass nor on the metallicity. (cf. )
    
    The mass of the elements returned to the gas phase by a star over its entire lifetime is accounted when it dies and the amount of the ejection is given by the stellar yield table according to the stellar initial mass and metallicity. There are a limited number of stellar initial metallicity available in the yield table (adopted from the NuPyCEE program, see footnote 12), i.e., Z=0.0127, 0.008, 0.004, and 0.0004 as is shown in Fig.~\ref{fig:steller_yields} below. We interpolate the yield as a function of stellar mass for the same initial metallicity but apply the yield with the closest initial Z without interpolation. I.e., the yield of a star with given mass switches according to its metallicity at Z=0.01035, 0.006, and 0.0022. This treatment is reasonable as the applied stellar yields have large systematic uncertainties. A more complicated interpolation treatment would not necessarily increase the reliability of the results. However, it does make the slope of the chemical evolution plots discontinuous at these switch values as highlighted by the dash-dotted lines in Fig.~\ref{fig:Z_over_X_time}.
    
    We note the difference between the "net yield", which is the mass of an element added or destroyed by a star from the gas phase, and the "total yield", which is the mass of ejected elements. The total yield is always positive but the net yield can be negative. We adopt the total yield table. The element ejection in our code depends only on the stellar initial mass and metallicity but not on its metal composition. Such an approximation is only valid for the stars with Solar composition and leads to errors in general situations. Simply applying the net yield cannot solve this problem as the difficulty lies in there being no yield table using stellar models with extremely non-solar initial abundances (especially for helium, see Section~\ref{sec:result} below) that can be applied for the massive elliptical galaxies.
    
    The mass of the most abundant elements H, He, C, N, O, Ne, Mg, Si, S, Ca, Fe, as well as the sum of all the elements heavier than helium are traced separately\footnote{It is possible to consider more elements or isotopes for relevant purposes in future studies.}. For low- and intermediate-mass stars, we apply the AGB yields given by \citet{2001A&A...370..194M}. For massive stars, we apply the yield given by \citet{1998A&A...334..505P}. The metallicity, [Z] (defined in Eq.~\ref{eq:Z}), helium mass fraction, Y, and $\alpha$-enhancement, [Mg/Fe], of the ejected gas from a single dying star (or supernova event) as a function of the stellar initial mass and metallicity are shown in Fig.~\ref{fig:steller_yields}. Massive stars exhibit higher yields and $\alpha$-enhancement while almost all stars generate ejections with a higher helium fraction than the Sun.
    \begin{figure}
        \centering
        \includegraphics[width=\hsize]{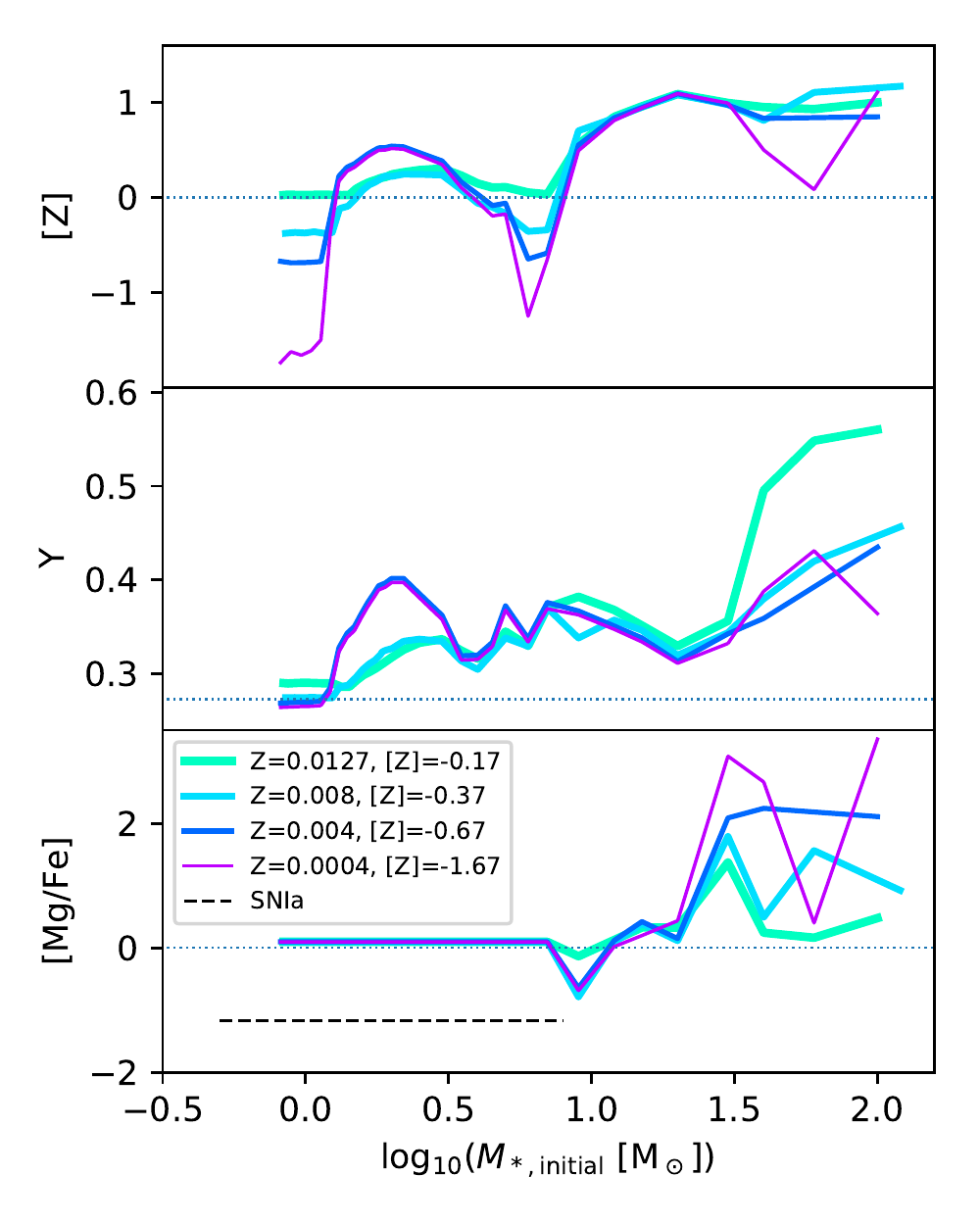}
        \caption{The metallicity, [Z] (as defined in Eq.~\ref{eq:Z}), helium mass fraction, Y, and $\alpha$-enhancement, [Mg/Fe], of the ejected gas from a dying star (or supernova event) as a function of the stellar initial metal mass fraction, Z, and the stellar initial mass, $M_{*,\rm initial}$, (or the initial mass of the possible SNIa progenitor, i.e., 3 to 8 M$_\odot$ shown by the dashed line). The thin dotted lines indicate the Solar value (The Solar value of Y=0.273 is adopted from \citealt{2010ApJ...719..865S}). The relation is interpolated from the stellar total yield table given by \cite{2001A&A...370..194M} and \cite{1998A&A...334..505P} where a Solar value stellar initial helium abundance is assumed for the stellar evolution model. Note the values are different from the original paper because of the difference between the "net yield" and the "total yield" (see text). The color code is the same as in Fig.~\ref{fig:1000_IMF_evolution}, and \ref{fig:extended_1000_IMF_evolution}.}
        \label{fig:steller_yields}
    \end{figure}
    
    The total metal ejection of dying stars, $M_{\rm metal}$, is calculated with $M_{\rm metal}=M_{\rm ini}-M_{\rm rem}-M_{\rm ej,H}-M_{\rm ej,He}$, with the symbols on the right-hand side being the stellar initial mass, final remnant mass\footnote{The original remnant mass given by the stellar yield table instead of the spline fitted one.}, ejected hydrogen mass, and ejected helium mass, respectively, instead of adding up the ejected mass of every element ($M_{\rm metal}=M_{\rm ej,Li}+M_{\rm ej,Be}+...$) since not all the elements are computed in the simulation.

    The metal evolution is traced by three metal indicators: (i) the gas-phase metallicity, i.e., the mass fraction of metals in gas, (ii) the stellar mass-weighted metallicity, and (iii) the approximated stellar luminosity-weighted metallicity. Element gravitational settling in the star is not considered and the stellar initial metallicity is adopted for the calculations as it has a minor effect (see Section~\ref{sec:solar_metal} below).
    
    The calculation of stellar luminosity-weighted results are as follows.
    First we consider the luminosity of a single star which is simplified by adopting only the main-sequence luminosity and there is no dependence on the stellar metallicity. Note that the luminosity variation during the AGB phase may not have a significant influence on the result since the average element abundance of the AGB stars should be a representation of the entire stellar population. Following \citet[their chapter 1.3.8]{2004adas.book.....D} for the low-mass stars and \citet[their chapter 5.7]{2006essp.book.....S} for the stars more massive than about 2 M$_\odot$, the empirically determined mass-luminosity relation is approximated in our calculation to be
    \begin{equation}\label{eq:stellar mass-luminosity relation}
    \frac{L}{L_\odot} =
    \begin{cases} 
    0.23\left(\frac{M_{*,\rm initial}}{M_\odot}\right)^{2.3}, & \frac{M_{*,\rm initial}}{M_\odot} < 0.421; \\
    \left(\frac{M_{*,\rm initial}}{M_\odot}\right)^{4}, & 0.421 \geqslant \frac{M_{*,\rm initial}}{M_\odot}<1.96; \\
    1.4\left(\frac{M_{*,\rm initial}}{M_\odot}\right)^{3.5}, & 1.96\geqslant \frac{M_{*,\rm initial}}{M_\odot}<55.41; \\
    32000\frac{M_{*,\rm initial}}{M_\odot}, & 55.41\geqslant \frac{M_{*,\rm initial}}{M_\odot},
    \end{cases}
    \end{equation}
    where $M_{*,\rm initial}$ is the stellar initial mass (same as the $M_{*,\rm initial}$ in Fig.~\ref{fig:stellar_lifetime_final_mass} and \ref{fig:steller_yields}). While the detailed shape of the mass-luminosity relation is important for understanding the shape of the stellar luminosity function \citep{1990MNRAS.244...76K,1993MNRAS.262..545K,2013pss5.book..115K}, the approximation used here suffices for the present purpose. Then the luminosity-weighted metallicity, Z$_{\rm lw}$, of all the living stars at time $t_n$ is calculated by:
    \begin{equation}
    Z_{\rm lw}(t_n) = \frac{\sum\limits_{t=t_1}^{t_n} \left[ \int_{m=m_{\rm min}}^{m_{\rm max}(t_n-t)} \xi(t,M_{*,\rm initial})L(M_{*,\rm initial})dm \cdot Z(t) \right]}{\sum\limits_{t=t_1}^{t_n} \left[ \int_{m=m_{\rm min}}^{m_{\rm max}(t_n-t)} \xi(t,M_{*,\rm initial})L(M_{*,\rm initial})dm \right]},
    \end{equation}
    where $\xi(t)$ is the gwIMF for the stellar population formed at time $t$ (see $\xi_{\rm IGIMF}$ in \citealt{2017A&A...607A.126Y} for definition), $Z(t)$ is the metallicity (or mass fraction for a given element) of the stellar population formed at time $t$, $m_{\rm max}$ is the mass of the most massive star that is still living at time $t_n$ for the stellar population formed at time $t$, and $m_{\rm min}$ is the lowest possible stellar mass. The luminosity-weighted element ratio, e.g., [Fe/H]$_{\rm lw}$ is calculated by a similar equation given in Appendix~\ref{sec:Luminosity-weighted element ratio}.
    
    \subsection{Modeling the rate of type Ia supernova}\label{sec:SNIa}
    
    SNIa are generally considered to be generated by binary systems containing degenerate stellar remnants \citep{2012PASA...29..447M}. As a consequence, the SNIa rate should depend on the spatial density of the potential progenitors of the binary stellar system, i.e., the dynamical encounter rate in stsar cluster \citep{2002ApJ...571..830S}. Since we do not have spacial modeling in our simulation, we simplify the consideration with the following two assumptions: 
    
    \begin{enumerate}
        \item We assume that the stellar density is similar for different stellar systems such that the rate of SNIa depends only on the star formation rate and the Delay Time Distribution (DTD) function but not on the density of the stellar system. This is not a good (see Section~\ref{sec:uncertainty SNIa} below) but commonly implied assumption in galaxy evolution models\footnote{This is the case not only for all the studies applying a DTD function, e.g., \citet{2013MNRAS.435.3500Y,2014MNRAS.445..970D}, but also those considering a specific SNIa formation model, e.g., \citet{2004MNRAS.347..968P}.}.

        \item We assume that the SNIa progenitors and their companion stars are born in the same star cluster. This is reasonable because, although stars and white dwarfs do leave their birth star cluster, the chance of them to meet and form a binary system outside their birth star cluster can be neglected. Under this consideration, the SNIa rate of a stellar population formed in a star formation epoch, $\delta t$, is not correlated with any former or later star formation epoch.
    \end{enumerate}
    
    Given the above assumptions, the standard number of SNIa for a canonical IMF is calculated individually for each stellar population according to its age. 
    
    The estimation of the SNIa rate that begins with a theoretical consideration of a specific SNIa formation model, and thus the SNIa progenitor number, suffers from the fact that the nature of the companion star is still under discussion (but see \citealt{2019arXiv190305115T}). Modern galaxy evolution studies avoid this problem by applying the DTD function that is observationally constrained \citep{2005A&A...441.1055G}. Our contribution applies such a recently constrained DTD function but deals simultaneously with the variable gwIMF that affects the SNIa number by re-normalizing the result.
    
    The DTD function of the SNIa is assumed to be a power-law following \citet{2012PASA...29..447M}:
    \begin{equation}\label{eq:DTD}
    N_{\rm SNIa,<t}(t) =
    \begin{cases} 
    0, & t \leqslant 40~{\rm Myr}, \\
    k \cdot t^{-1}, & t>40~{\rm Myr},
    \end{cases}
    \end{equation}
    where $N_{\rm SNIa,<t}(t)$ is the number of exploded SNIa for a stellar population with age $t$, and $k$ is a normalization parameter such that the DTD integral over 10 Gyr is $2.25\cdot 10^{-3} \rm M_\odot^{-1}$, i.e., about 2 SNIa for every 1000 M$_\odot$ of stars formed. The value $2.25\cdot 10^{-3} \rm M_\odot^{-1}$ is suggested by observation and has an error of roughly about $\pm 1\cdot 10^{-3} \rm M_\odot^{-1}$ \citep{2012PASA...29..447M}, where the error is not considered in our code. 
    
    Since we are dealing with a non-canonical IMF and the number of SNIa depends on the number of potential SNIa progenitors, that is, stars with initial mass between about 1.5 and 8 M$_\odot$\footnote{It is commonly assumed that the maximum stellar mass able to produce a degenerate C/O white dwarf is 8 M$_\odot$ \citep{1983A&A...118..217G}. The minimum possible binary mass ensuring that the smallest possible white dwarf can accrete enough mass from the secondary star to reach the Chandrasekhar mass is about 2 or 3 M$_\odot$. Here we assume that all stars with a initial mass from 1.5 M$_\odot$ to 8 M$_\odot$ has the same change to be part of a SNIa event. It may not be the best approximation (cf. \citealt{2006ApJ...648..230L}) but the chemical evolution results of massive galaxies are not sensitive to the choice of 1.5 M$_\odot$ limit (see footnote 19 below).}, $N_{\rm SNIa,<t}(t)$ needs to be re-normalized accordingly with the new number of potential SNIa progenitors per stellar mass formed times the possibility of having its companion also in the correct mass range:
    \begin{equation}\label{eq:renorm}
    \begin{split}
    N_{\rm SNIa,<t,nor}(t) &= N_{\rm SNIa,<t}(t)\cdot \frac{p_{\rm gwIMF}}{p_{\rm diet-Salpeter}}, \\
    p &= \frac{n_{1.5,8}}{M_{0.08,150}} \cdot \frac{n_{1.5,8}}{n_{0.08,150}},
    \end{split}
    \end{equation}
    where $N_{\rm SNIa,<t,nor}(t)$ is the re-normalized number of SNIa, $p$ is the normalization parameter depending on the assumed IMF. Thus the number of SNIa is adjusted by the $p$ value assuming any given gwIMF over the $p$ value assuming the "diet-Salpeter IMF", as defined in \citealt{2001ApJ...550..212B} (differs from the Salpeter-IMF by having a flat slope below 0.35 M$_\odot$), is the assumed gwIMF for the empirical measurements of the SNIa rate. Parameter $n$ is the number of stars within the mass range given by its subscript (in the unit of M$_\odot$), and $M$ is the corresponding total mass of the stars in the given mass range. 0.08 M$_\odot$ and 150 M$_\odot$ are the assumed lowest and highest possible stellar mass in our model following \cite{2017A&A...607A.126Y}. Thus the normalization parameter $p$ stands for the number of stars in the right mass range of being the primary star in the progenitor system of SNIa for a stellar population with known total mass, times the possibility that its companion star is also in the right mass range\footnote{The correct mass range of the companion star should depend on the primary star. But since the SNIa progenitor model is still uncertain and other uncertainties of the SNIa model are larger, a more detailed consideration does not improve the accuracy of our estimation. In addition, the average level of the SNIa rate is determined mostly by the empirical parameter $k$ in Eq.~\ref{eq:DTD} while the re-normalization procedure is only used as a secondary adjustment to compensate for the non-canonical IMF effect instead of trying to calculate the SNIa rate directly. Thus the re-normalized SNIa rate is not sensitive to the exact mass boundary we chose in Eq.~\ref{eq:renorm} as long as the gwIMF is not top-light. This is confirmed when we apply a different lower boundary of 3 M$_\odot$. Thus it is reasonable to apply the same mass range of 1.5 to 8 M$_\odot$ in Eq.~\ref{eq:renorm} for both the primary and secondary star to reduce the computational cost.}. Note that the possibility of any star being in a binary system is absent in Eq.~\ref{eq:renorm} as this possibility will be canceled in the term $p_{\rm gwIMF}/p_{\rm diet-Salpeter}$ according to the assumption 1 in Section~\ref{sec:SNIa}. Considering the possible variation of the stellar dynamical encounter rate, and thus of the fraction of hard binaries, it may be possible that relatively more SNIa occur at high-SFRs and thus in massive galaxies.
    
    \subsection{The definition of the metallicity}\label{sec:definition_metallicity}
    
    Through out the study we adopt an earlier Solar metal\footnote{The "metal" element includes all the elements heavier than helium.} mass fraction measurement of $\rm Z_\odot=0.01886$ from \cite{1989GeCoA..53..197A} such that it is easier to compare our study with earlier galaxy evolution simulations (see Section~\ref{sec:solar_metal} below). Thus the metallicity is defined as:
    \begin{equation}\label{eq:Z}
    \rm [Z]=log_{10}(Z/0.01886),
    \end{equation}
    where Z is the metal mass fraction of the target.
    
    When comparing the relative metal and hydrogen abundances, this paper uses [Z/X], where X is the hydrogen mass fraction of the target, instead of [Z/H] to highlight the definition of this symbol, being:
    \begin{equation}\label{eq:Z_over_X}
    \rm [Z/X] = log_{10}(Z/X) - log_{10}(Z_\odot/X_\odot),
    \end{equation}
    where X$_\odot$=0.70683 is the Solar mass fraction of hydrogen adopted from \cite{1989GeCoA..53..197A}. [Z] and [Z/X] gradually deviate from each other as stellar helium abundance increase to super-Solar values.
    
    On the other hand, the symbols written with the name of the element, e.g., [Fe/H] and [Mg/Fe], follow the canonical definition calculated by the atom number ratio.
    
\section{Uncertainties}\label{sec:uncertainty}
    
    The different input components that a galaxy chemical evolution simulation is most sensitive to and which are not well constrained are briefly reviewed to get an idea of the reliability of our simulation. The following factors are able to significantly affect the galaxy chemical evolution via changing, e.g., the relative influence of SNIa and type II supernovae (SNII). The input parameters typically have an uncertainty larger than 0.2 decimal exponent (dex) in the logarithmic scale (that is an error of about 50\%). This section evaluates the influence of these uncertainties.
    
    \subsection{Gas flow and efficiency of star formation}\label{sec:gas_flow}
    
    Here we discuss whether the applied simplifications of instant gas mixing and no gas-flows are good approximations.
    
    \cite{2014A&A...566A..71L,2017A&A...606A..36C,2018ApJ...859..164B} find a lack of star-formation driven outflows in observed low-mass and low-/intermediate-redshift galaxies.
    Outflows in more massive elliptical galaxies would be more difficult since they have a deeper potential well, consistent with the suggestion from X-ray \citep{2012ASSL..378.....K} and chemical evolution studies \citep{2000MNRAS.316..786F}. 
    In principle, with the violent star formation activity of massive galaxies, the galaxy gas can be heated to a high temperature \citep{2001MNRAS.328..461O} which may escape the galaxy and enrich the intercluster medium (ICM) \citep{2014MNRAS.444.3581R,2017A&A...603A..80M} and/or be locked up in the hot corona around the galaxy, thereby having a significantly long cooling time and thus not being available for further recycling \citep{2016MNRAS.461.1760H}.
    A positive correlation between the gas velocity and the galaxy-wide SFR is indeed evident in observations \citep{2017ApJ...850...51S,2019arXiv190403106S} but it is unclear if such an outflowing gas is a galactic fountain, that can be recycled, or galactic wind, that cannot (see \citealt{2012ceg..book.....M} for definition). 
    
    For the inflow, since the star formation and galaxy assembly of the massive elliptical galaxies are accomplished within a Gyr timescale \citep{2006ARA&A..44..141R}, their formation can be approximated as an initial rapid collapse during which gas inflow from afar would not have played a significant role. 
    
    Thus we discuss the hydrodynamic considerations employed by other simulation studies and their possible effects:
    
    \begin{itemize}
    
    \item The primordial gas inflow regulates when and how much gas is accreted to the galaxy and thus the SFH. An exponentially decreasing gas-inflow rate is often assumed in semi-analytical simulations to simulate the SFH of the starburst (see for example \citealt{2009A&A...504..373C,2011A&A...530A..98P,2015MNRAS.449.1327V}). Our model achieves a similar SFH by specifying it.
    
    The amount of gas infall is usually used as a free parameter to adjust the final metallicity of a galaxy\footnote{Note that the infall model increase the mental production of a galaxy and ends up with less metal-poor stars thus is a solution of the G-dwarf problem \citep{2001PASP..113..137P,2009nceg.book.....P,2012ceg..book.....M}.}. Our model does not have gas-flows and use the initial gas mass as a free parameter to adjusts the final metallicity.
    
    \item The galactic fountain refers to the gas flows within the galaxy. The general effect is to delay the gas mixing between the SN ejecta and the ISM thus the recycling of the elements, which appears to have a limited effect on the chemical evolution of disk galaxies \citep{2009A&A...504...87S}. For the simulation of high mass ellipticals that have a high SFR and supernova explosion rate (see Fig.~\ref{fig:SN_number_evolution} below), gas-mixing should be more efficient and we consider the instantaneous mixing of gas is a good approximation.
    
    \item A galactic wind can be included in our model. It would move the enriched gas out of the galaxy thereby reducing the final total gas mass and metallicity of the simulated galaxy.
    
    A uniform galactic wind, where element ratios of the ejected gas are the same as in the well-mixed gas phase, is often applied in galaxy evolution studies\footnote{Note that the lockup of gas by low-mass stars has the same effect as a uniform galactic wind. The mass fraction in brown dwarfs and planets is smaller than about 4\% \citep{2013pss5.book..115K} thus has a negligible effect and will not be considered.}. For massive ellipticals, we test with our model that if the mass of the galactic wind is the same as the mass of stars being formed at the time, as suggested for dwarf irregular galaxies \citep{2012ceg..book.....M}, the metal abundance of the galaxy at $t=13$ Gyr would decrease by about 0.1 dex, and the metal abundance ratios change less than 0.05 dex. A significant influence on the simulated chemical evolution results would require a strong galactic wind scenario with an outflow-mass--star-formation-mass ratio much higher than 1. According to \cite{2017ApJ...850...51S,2019arXiv190403106S}, the estimated total mass of outflows for high-redshift galaxies should be less than about 5 times the total stellar mass. This amount of outflow would not affect significantly our simulation results even if all the outflow becomes a galactic wind. The effect of galactic wind may not be significant either for lower mass galaxies as suggested by \citet{2007MNRAS.375..673K}. Since there is no direct observational evidence supporting the strong galactic wind scenario, we decide to demonstrate only the simulations without a galactic wind.
    
    Other than the uniform galactic wind, the wind may contain more metals (i.e., a selective outflow, \citealt{2009nceg.book.....P}), as metal elements experience larger radiation pressure \citep{2008A&A...489..555R}; or the wind may be largely composed of the $\alpha$-element-rich supernova ejections \citep{2018MNRAS.474.1143L} probably under strong starburst circumstances \citep{2013MNRAS.435.3500Y}. The enriched wind taking away the metals more efficiently and selectively is used to explain anomalous observations \citep[their equation 2.36]{2012ceg..book.....M}. A time-dependent galactic wind might be necessary to explain the metal abundance of gas and stars together as suggested by \citet{2018MNRAS.474.1143L}. These selective treatments have large uncertainties and are not applied in our model. They will result in a lower metallicity and [$\alpha$/Fe] if applied.
    
    \item The efficiency of star formation (the parameter $\nu$ in unit of Gyr$^{-1}$ as defined in \citealt{1994A&A...288...57M}) strongly affects the metallicity. It depends on baryonic physics that influences the gas recycling process such as gas expulsion, mixing, and cooling processes. In some studies, the efficiency of star formation is assumed to be a constant for galaxies with order of magnitude differences in mass or SFR. In others, especially those trying to reproduce the downsizing of the star-formation timescale, the efficiency of star formation is assumed to be higher in more massive galaxies \citep{1994A&A...288...57M,2009A&A...504..373C}. The uncertainty of the efficiency of star formation is about 0.2 dex and leads to an uncertainty in the chemical evolution simulation results larger than 0.2 dex. Since we aim to apply a fixed SFH and vary only the IMF assumption, our model does not involve the parameter $\nu$. Instead, we use the parameter $f_{st}$ (Table~\ref{tab:abbreviation}) to specify how much stars are formed relative to the initial gas mass.
    
    \end{itemize}
    
    We emphasize that the necessity of the gas-flow model depends on the galaxy type and assumptions about the IMF. Gas flows were initially introduced to explain the metal distribution of the stars in the Solar neighborhood and the Galactic disk, e.g., the G-dwarf problem of the Milky Way. For an elliptical galaxy with a variable gwIMF, we need to re-evaluate the situation. Only when the simpler mode without gas-flows fails should we introduce more free parameters to describe the gas-flow (similar to what \citealt{2000A&A...353..269M} explore for disk galaxies).
    
    As stressed in \citet[their section 11.2.1]{2009nceg.book.....P}, a constraint on any modeling parameter requires an independent observable. 
    This is difficult when we consider distant elliptical galaxies that cannot be resolved into single stars and are likely to have a non-canonical TIgwIMF\footnote{Additional spectral line features do not give independent constraints since they involve (i) multiple element abundances and (ii) the surface temperature of stars with different masses (thus the TIgwIMF), both being uncertain if the IMF is variable.}. Without proper constrains, the gas in- and out-flow models are not reliable and only complicate the analysis.
    
    Essentially, the formation of elliptical galaxies occurs on a much shorter time-scale than the disk galaxies that apart from the initial infall the majority of star-formation is finished before additional gas accretion or galaxy wind can play a role (cf. \citealt{1999MNRAS.302..537T} their "fast clumpy collapse model"). Models without gas-flow are thus likely good and in particular computationally efficient approximations to describe the formation and evolution of elliptical galaxies (especially for the central region where most of the star formation and metal enrichment happens, cf. \citealt{2019A&A...625A..11A}).
    
    \subsection{Initial gas metallicity}\label{sec:Initial gas metallicity}
    
    Most massive elliptical galaxies are considered to be formed at the earliest time \citep{2006ARA&A..44..141R,2010MNRAS.404.1775T,2019ApJ...870..133E} when the gas is not pre-enriched. The effect of pre-enrichment, if it occurred, is to eliminate the very metal-poor stars and may raise the final metallicity of the galaxy. 
    
    In our simulation, the metallicity dependence of the gwIMF leads to a feedback effect since a low-metallicity environment results in a more top-heavy gwIMF \citep[their Fig. B.1]{2017A&A...607A.126Y} leading to a higher IMF-weighted metal yield \citep{2015MNRAS.446.4168R}. This effect makes the final metallicity of the galaxy not sensitive to its initial value.
    
    The initial gas metallicity, as well as the metal-dependence of the IGIMF formulation, mainly affects the property of the very first generations of stars polluted by massive star ejections, which is out of the scope of the current study. Later stages of the chemical evolution depend far more on the details of the SFH and IMF slope than on the initial metallicity.

    \subsection{Type II supernovae}
    
    The stellar evolution models give different SNII yields. \citet{1998MNRAS.296..119T,2005A&A...436..879K,2018ApJ...861...40P} shows that the IMF averaged [Mg/Fe] yield between two table may differ as much as 0.3 dex. This uncertainty is inherited by the final metallicity of the galaxy and modifies the element abundance ratios \citep[their section 4.2]{1998MNRAS.296..119T,2017ApJ...835..224A}. Consider the rotation of pre-supernova star in the stellar evolution model gives even larger yield uncertainties, especially for isotopes \citep{2005A&A...433.1013H,2018A&A...618A.133C}.
    
    In addition, not all the high mass stars will end up as a SNII. Direct black hole formation without supernova \citep{2015PASA...32...16S} and different types of core-collapse supernovae due to initial mass, metallicity, and angular momentum differences of the supernova progenitors \citep{2003ApJ...591..288H,2016ApJ...821...38S}, kilonovae (supposedly two neutron stars or a neutron star and a black hole merger events), as well as other cosmic selection effects that may modify the SNII rate \citep{2011ApJ...738..154H} or favour contributions from supernovae in a certain mass range or favour the recycling of certain elements \citep{2005A&A...436..879K} are currently not considered by most galaxy evolution simulations.
    
    Due to these concerns, the yield of a stellar population is uncertain even with a known IMF.
    
    \subsection{Type Ia supernovae}\label{sec:uncertainty SNIa}
    
    The SNIa yields also show large uncertainties. For example, the ejecta from a single SNIa event have a [O/Mg] ranging in different references from about -0.43 in \citet{1984ApJ...286..644N,2013MNRAS.429.1156S} and about -0.25 in \citet{1986A&A...158...17T,2003fthp.conf..331T} to about 0.05 in \citet[this is our applied SNIa yield]{1997MNRAS.290..623G}; \citet{1999ApJS..125..439I}. Thus applying a different SNIa yield table and consider different $\alpha$ element certainly affect the resulting [$\alpha$/Fe]. Note that dying stars give a [O/Mg] yield ranging from about 0 dex to as high as 1.6 dex depending on the stellar initial mass and metallicity, which leads to an oxygen overproduction relative to magnesium in galaxy evolution simulation relative to observations. Choosing a higher SNIa Mg yield could in principle be helpful (cf. \citealt{2004A&A...421..613F,2011A&A...530A..98P}) but such discussion is out of the scope of the current publication. The ejected iron mass shows more consistency between different references, typically being $0.663\pm 0.066$ M$_\odot$ per SNIa event. The applied Fe yield in the current simulations is 0.744 M$_\odot$ per SNIa event \citep{1997MNRAS.290..623G}.
    
    A different shape of the SNIa DTD can have a significant effect on the abundance ratio between the $\alpha$ elements and iron peak elements of a galaxy. A single power law, a broken power law, and a bivariate DTD have been proposed \citep{2006MNRAS.370..773M,2009A&A...501..531M,2010ApJ...722.1879M,2013MNRAS.435.3500Y}. The estimated power-index, assuming a single power-law shape, ranges from 0.9 to 1.3. The time delay between the formation of a single age stellar population and the peak of its SNIa rate (the 40 Myr in Eq.~\ref{eq:DTD}) is also uncertain by several dozens of Myr. However, it does not have a notable affect on the SNIa history of a galaxy where SNIa rate peaks at a Gyr timescale since the SNIa explosion history is a convolution of Eq.~\ref{eq:DTD} with the Gyr timescale SFH. Thus the uncertainty of the 40 Myr applied in Eq.~\ref{eq:DTD} can be ignored for the present purpose.
    
    The largest uncertainty introduced by Eq.~\ref{eq:DTD} to the galaxy chemical evolution is from the normalization parameter, $k=2.25\cdot 10^{-3} \rm M_\odot^{-1}$. The empirical measurements of the parameter $k$ have not only a large scatter of about $\pm1$ SNIa per 1000 M$_\odot$, but also tensions between different measures. Studies of the iron abundance in the ICM have shown that the parameter $k$ should be higher than 3.4 \citep{2010ApJ...722.1879M}\footnote{\citet{2010ApJ...722.1879M} assumes a diet Salpeter IMF. The conclusion for other Milky-Way-like IMFs will be similar. The conclusion for a top-heavy IMF would require an even larger normalization parameter as the relative number of SNIa precursors is reduced in a top-heavy IMF.}, being inconsistent with other direct measurement methods. This suggests a systematic error possibly relating to the intrinsic variation due to environmental differences. 
    
    \subsection{Recent star formation activity}\label{sec:Recent_star_formation}
    
    Later star formation after the primary star burst is likely \citep{2018ApJ...861...13C,2018arXiv181206980M}. A SFH with a long tail of low-SFR after the starburst, or a later second smaller starburst that contributes a decent portion (10\%, e.g., \citealt{2013MNRAS.432..359T}) of stars formed from metal-enriched gas may not influence the stellar mass-weighted metallicity but can alter the luminosity--weighted metallicity significantly as is show in Appendix~\ref{sec:lognorm_results}. The luminosity-weighted metal abundances are closer to the metal abundance of the gas-phase that is higher than the stellar mass-weighted metallicity. The different SFH of galaxies and especially the difference in recent star formation activity will introduce an intrinsic scatter on the interpreted galaxy properties.
    
    \subsection{The adopted value for the Solar metallicity}\label{sec:solar_metal}
    
    The Solar metal mass fraction, $\rm Z_\odot$, i.e., the mass fraction of the elements heavier than helium, have been declining continuously over the past decades, from $1.89\%$ in \cite{1989GeCoA..53..197A} to about $1.34\%$ in \cite{2009ARA&A..47..481A}. Thus the application of a different Solar metallicity can lead to a 0.15 dex difference in the resulting [Z]. However, many observational papers do not explicitly define the Solar metallicity they applied. Following \cite{1995ApJS..101..181W}, this paper adopts the earlier Solar metallicity measurement of $\rm Z_\odot=0.01886$ from \cite{1989GeCoA..53..197A} as mentioned in Section~\ref{sec:definition_metallicity}.
    We avoid the involvement of the Solar metallicity when adopting the stellar evolution yield table by applying Z instead of [Z]. Finally, the present-day photosphere metallicity of the Sun is similar to its bulk or protosolar metallicity with a difference in $\rm log_{10}(Z)$ smaller than 0.03 dex according to \citet{2009ARA&A..47..481A}. Thus we do not distinguish the two as other parameters introduce far larger uncertainties.

\section{Simulation results}\label{sec:result}
    
    In this section, the effect of applying different IMF assumptions is shown. Two simulations with the exact-same set up (e.g., SFH and initial gas mass) except for the assumption on the gwIMF are carried out, one assuming the gwIMF is the invariant canonical two-part power-law IMF \citep{2001MNRAS.322..231K}, the other assuming the gwIMF is metallicity and SFR dependent according to the latest IGIMF theory formulation \citep[their IGIMF3 model]{2018A&A...620A..39J}. The $f_{st}$ is set to be 100\%, i.e., the sum of the initial stellar masses of all the stars ever formed throughout the galaxy formation history is the same as the initial gas mass\footnote{The highest possible $f_{st}$ value for a stellar population with the canonical gwIMF is roughly 150\% since the stellar population returns about 1/3 of its mass to the gas-phase after 100 Myr (this fraction increase to about 1/2 after 10 Gyr).}. The boxy SFH in Fig.~\ref{fig:SFH} is assumed, where the SFR is fixed at 1000 M$_\odot$/yr and lasts for $t_{sf}=1$ Gyr. We simulate the chemical evolution of a galaxy for 13 Gyr (the nominal age of the Universe) and result in the stellar-mass-weighted, in the gas-mass-weighted, as well as a stellar-luminosity-weighted (approximated, see Section~\ref{sec:Metal_enrichment}) element abundance evolution, the evolution of total gas, living star, and stellar remnant mass, the evolution of the TIgwIMF, and the evolution of SN rates.
    
    In addition, we compare the galaxy evolution assuming different SFHs (the boxy and log-normal SFH shown in Fig.~\ref{fig:SFH}) under the IGIMF theory. The results are shown in Appendix~\ref{sec:lognorm_results}.
    
    \subsection{The evolution of the TIgwIMF}
    
\begin{figure}
        \centering
        \includegraphics[width=\hsize]{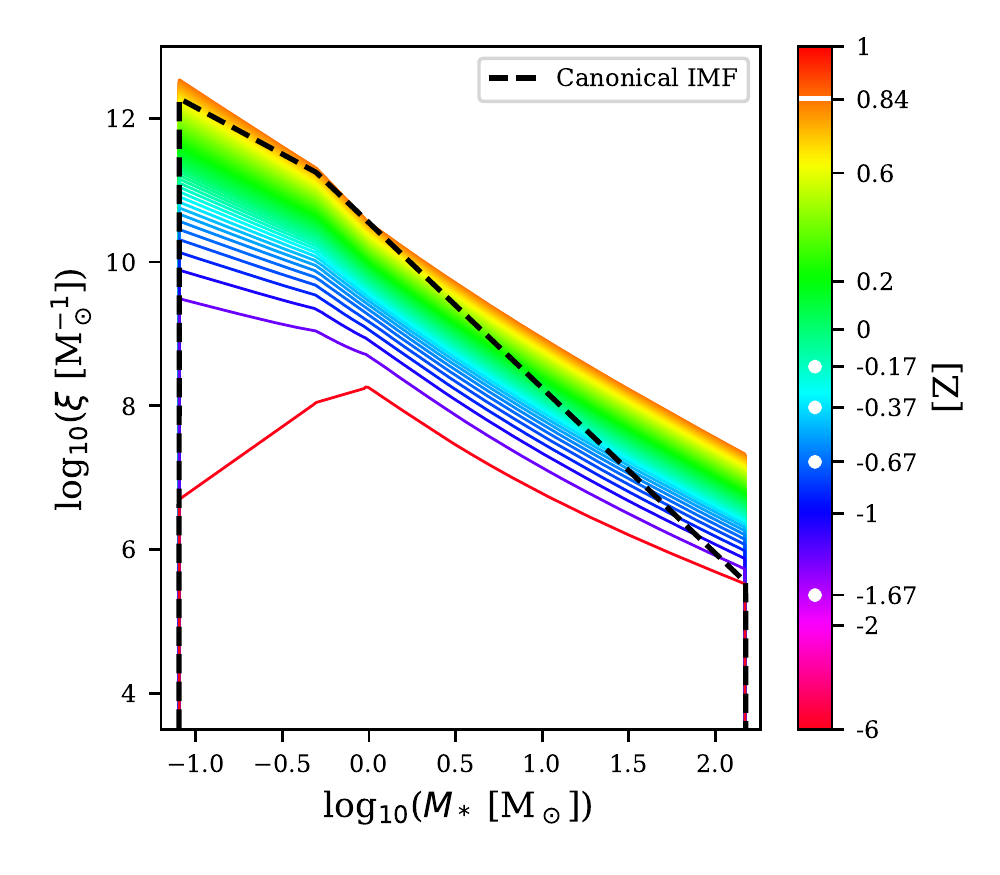}
        \caption{The evolution of the TIgwIMF for a galaxy with the boxy SFH shown in Fig.~\ref{fig:SFH}, where $\xi$ is the total number of stars within a unit mass range. The assumed lowest and highest possible stellar mass is 0.08 M$_\odot$ and 150 M$_\odot$, respectively, following \cite{2017A&A...607A.126Y}. The solid lines represent the TIgwIMF after 10, 20, ..., and 1000 Myr since the beginning of the star formation (from lowest to top most line, respectively). The color of the lines indicates the metallicity, [Z] (as defined in Eq.~\ref{eq:Z}), in gas at the time. The color code is the same as in Fig.~\ref{fig:stellar_lifetime_final_mass} and \ref{fig:steller_yields} where the white dots in the colorbar indicate the available initial metallicity values of the stellar yield table (see Section~\ref{sec:Metal_enrichment}) and the white line in the colorbar indicates the initial metallicity of the last stellar population, i.e., the most metal-rich population. The dashed line is the canonical IMF given by \cite{2001MNRAS.322..231K}, normalized to have the same $\xi$ as the final TIgwIMF at $M_*=1$.}
        \label{fig:1000_IMF_evolution}
    \end{figure}

    Under the IGIMF theory framework, the TIgwIMF changes gradually from top-heavy bottom-light to top-heavy bottom-heavy\footnote{Note that the gwIMF can become top-light at a late time if the SFR drops to a lower value but the TIgwIMF stays top-heavy.} as is shown in Fig.~\ref{fig:1000_IMF_evolution}, in conformity with the pioneering variable-IMF suggestion for producing the old and metal rich stellar population of ellipticals \citep[their section 4.2.1]{1996ApJS..106..307V}.
    
    Note that the TIgwIMF as shown in Fig.~\ref{fig:1000_IMF_evolution} combines all gwIMFs over the formation history of the galaxy, i.e., it comprises all stars ever formed. The here shown TIgwIMF does not resemble the gwIMF of the Milky Way that has a different SFH.
    
    The IMF is top-heavy due to the high SFR in the assumed starburst SFH. Consistent with \cite{2013ApJ...779....9B}, there is a lack of low-mass stars at early times when the environment is metal-poor; while the low-mass end evolves to a bottom-heavy form as the metallicity increases. The time-dependent TIgwIMF variation at the low-mass end may be relevant for the bottom-heavy TIgwIMF determined by some observations (see the generated bottom-heavy gwIMF and a more in-depth discussion in \citealt{2018A&A...620A..39J}).
    
    Note that the massive stars have a short lifetime and would not be observable in the present day. The observations on the still living low-mass stars do not give information about the high-mass IMF slope except for their observed metallicity which can be compared with the prediction given by the galaxy chemical evolution simulation assuming a certain IMF theory.
    
    \subsection{The gas, living star, and stellar remnant mass evolution}\label{sec:mass_evolution} 
    
\begin{figure}
        \centering
        \includegraphics[width=\hsize]{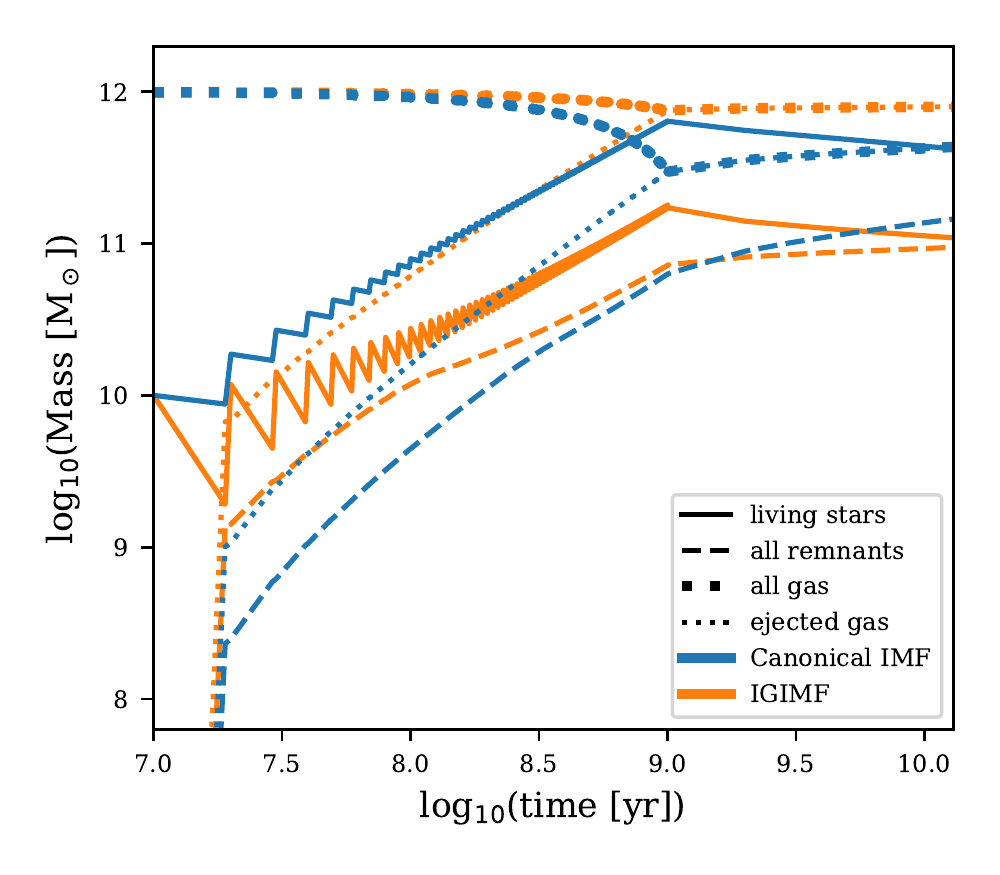}
        \caption{The mass evolution of gas, living stars, and stellar remnants as a function of time for the boxy SFH shown in Fig.~\ref{fig:SFH}. The orange and blue lines are, respectively, the results obtained assuming the gwIMF is given by the IGIMF theory and the invariant canonical IMF. The same color code is applied from Fig.~\ref{fig:mass_evolution} to \ref{fig:Mg_over_Fe_time}. The "all gas"includes gas outside the galaxy that is constrained by the galactic potential (see Section~\ref{sec:gas_flow} above). Different from Fig.~\ref{fig:1000_IMF_evolution}, here the simulated values at 10, 19, 20, 29... Myr are demonstrated, where star formation happens at 10, 20... Myr (see Section~\ref{sec:time_step}). The death of stars with a lifetime shorter than the star formation time step (10 Myr) causes the serrated shape of the lines.}
        \label{fig:mass_evolution}
    \end{figure}

    The difference in the mass evolution of gas, living stars, and stellar remnants as a function of time assuming the IGIMF theory and invariant canonical IMF are shown in Fig.~\ref{fig:mass_evolution}.
    
    Note that the total gas shown includes the circumgalactic medium that is trapped in the galactic potential. As our model assumes no galactic wind, all the gas is considered as trapped gas. The total gas mass is reduced less significantly in the IGIMF theory as the massive stars return most of their mass to the gas-phase quickly after their formation \citep[their figure 1]{2006ApJ...648..230L}. We consider that the star formation is not truncated due to a lack of gas but the fact that gas is heated to a high temperature, i.e., the apparent high gas mass in our simulation is the mass of a hot gas halo with a significantly long cooling time precluding it from forming new stars (cf. \citealt{2016MNRAS.461.1760H}).
    
    The stellar remnant mass is comparable to the mass of living stars in the IGIMF theory framework, while the canonical result is that the remnant mass is much smaller than the stellar mass. Thus the dynamical mass-to-light ratio of the studied massive elliptical galaxy would increase by about a factor of two under the IGIMF theory in comparison with applying the canonical IMF,
    in line with the IMF mismatch parameter of about 2 for massive ellipticals in most\footnote{Galaxies with a lower or normal dynamical mass-to-light ratio (i.e. with a IMF mismatch parameter close to one) can be the result of a different SFH and/or from mergers of less-massive E galaxies.} of the cases as evident in \citet{2017ApJ...838...77L,2018MNRAS.476..133O}.
    
    \subsection{The number of supernovae}

\begin{figure}
        \centering
        \includegraphics[width=\hsize]{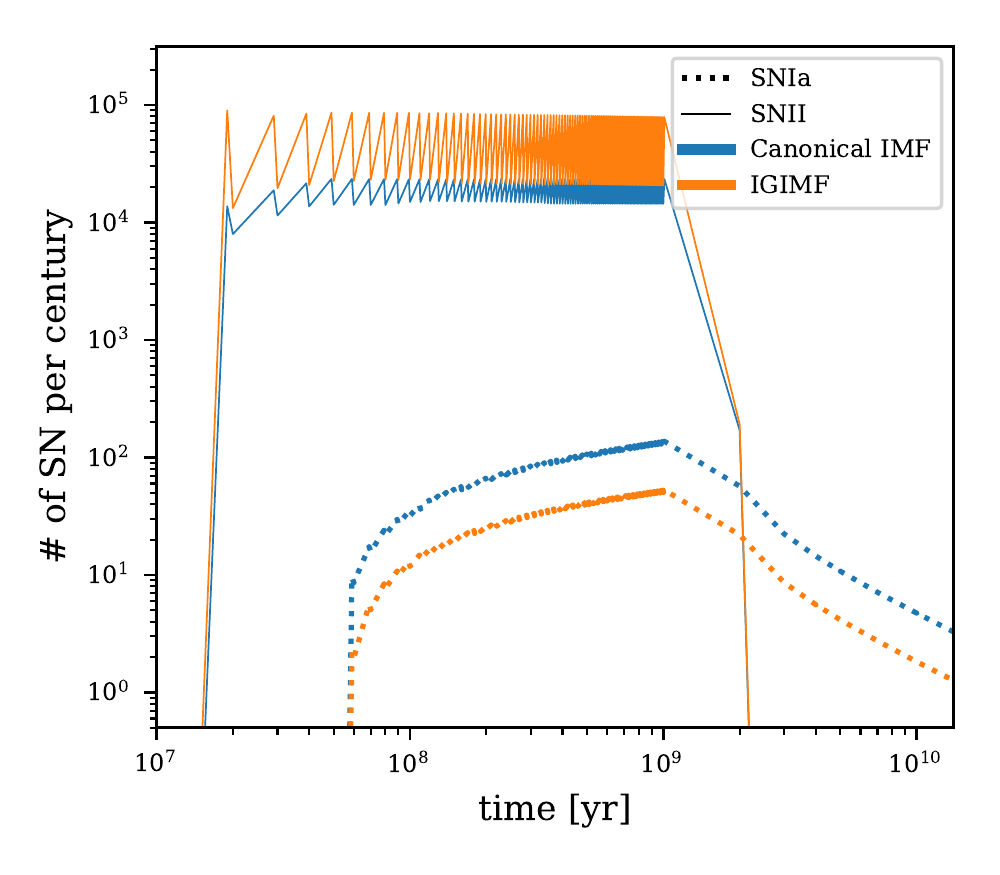}
        \caption{The number of SNIa and SNII per century as a function of time for the boxy SFH shown in Fig.~\ref{fig:SFH}. As in Fig.~\ref{fig:mass_evolution}, the orange lines are the results applying the IGIMF theory while the blue lines are applying the invariant canonical IMF. The thick lines below and the thin lines above are the rates for SNIa and SNII, respectively. The serrated shape is caused by the simplification that all the stars formed in a 10 Myr star formation epoch have an identical age.}
        \label{fig:SN_number_evolution}
    \end{figure}

    The time evolution of the number of SNIa and SNII adopting different IMF assumptions is plotted in Fig.~\ref{fig:SN_number_evolution}.
    
    The high SNII rate is a direct result of the assumed SFH. With the commonly assumed relation that more massive galaxies have a shorter $t_{sf}$ (see Section~\ref{sec:SFH} above), the starburst has to be extremely violent. Thus the galaxy produces a large number of SNII at early times with about 0.5 and 1 SNII event per day under the assumption of the invariant canonical IMF and the IGIMF theory, respectively (for a similar discussion in the context of the formation of ultra-compact dwarf galaxies see \citealt{2017A&A...608A..53J}). This seems to be consistent with the stronger emission lines observed for higher redshift galaxies \citep{2013ApJ...777L..19L,2015MNRAS.454.1393S}.
    
    On the other hand, the number of SNIa per stellar mass formed is reduced for the simulated high-SFR galaxy as the variation of the gwIMF changes the $p_{gwIMF}$ in Eq.~\ref{eq:renorm}. As is shown in Fig.~\ref{fig:SN_number_evolution}, the resulting SNIa rate applying the IGIMF theory is about one third of the simulation applying the canonical IMF (cf. \citealt{2010ApJ...716..384L}). Since SNIa produce more than half of the iron in galaxies, this suggests that the iron as a tracer may underestimate the metallicity (see Section~\ref{sec:Iron abundance} below).
    
    \subsection{Element abundances}\label{sec:Element_abundances}
    
    The mass-/luminosity-averaged [Z/X], Y, [Fe/H], and [Mg/Fe] evolution for stars and gas comparing different IMF assumptions are shown in Fig.~\ref{fig:Z_over_X_time} to Fig.~\ref{fig:Mg_over_Fe_time}. 
    
    Note that our gas-phase element abundances are mass averaged which may differ from the emission-weighted observations that depend on the local gas temperature and density. Note also that the observed stellar luminosity-weighted metallicity should be lower than our estimate which adopts the luminosity without the metallicity dependence (see Section~\ref{sec:Metal_enrichment} above) since metal-poor stars are more luminous.
    
    The behavior of the evolution for different elements shown in the following sections are similar. The stellar luminosity-weighted results are close to the gas-phase results during the star formation and quickly recover to the stellar mass-weighted results as the luminous massive stars die out on a hundred Myr or Gyr timescale.
    
    \subsubsection{Metallicity}
    
\begin{figure}
        \centering
        \includegraphics[width=\hsize]{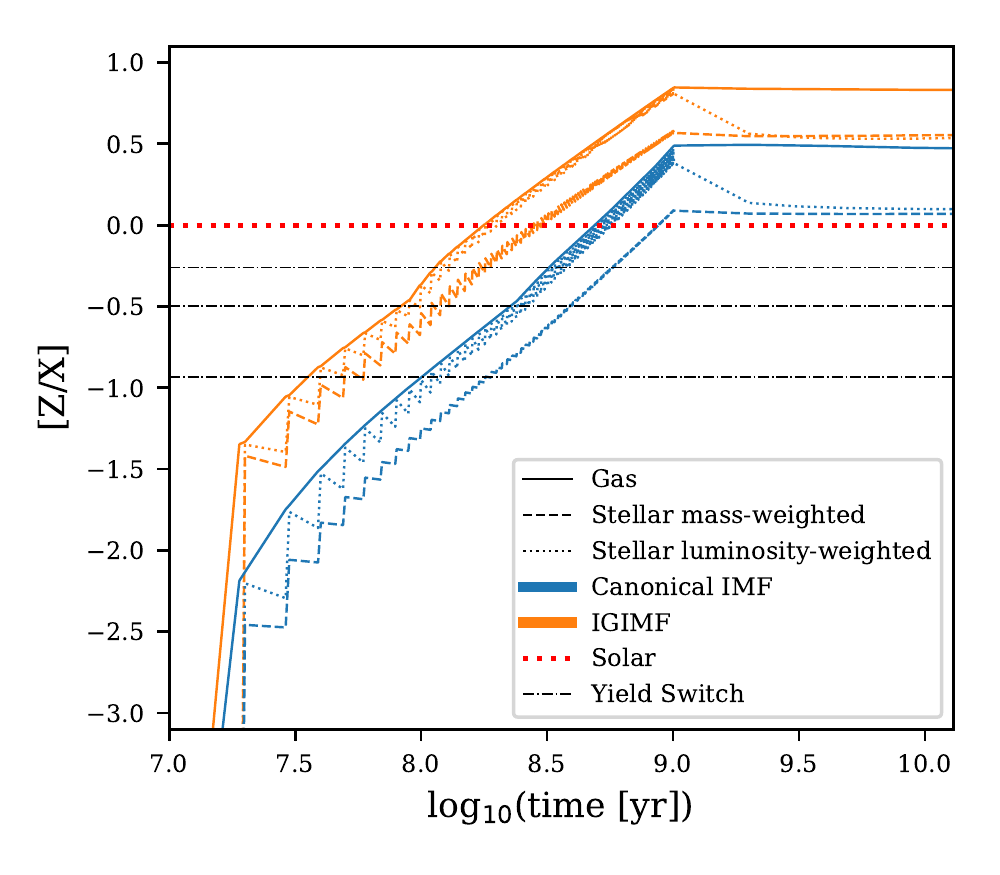}
        \caption{The evolution of gas and mass averaged stellar [Z/X] for the boxy SFH shown in Fig.~\ref{fig:SFH}, where [Z/X] is defined in Section~\ref{sec:solar_metal}. Simulations applying the IGIMF theory and fixed canonical IMF are shown as the orange and the blue lines, respectively. The solid, dashed, and dotted lines represent the value calculated for gas phase, mass average, and luminosity average (approximation, see text), respectively. The horizontal dash-dotted lines mark the boundaries where the stellar yield table is changed (at [Z/X]$\approx$[Z]=-0.26, -0.497, -0.933, see Section~\ref{sec:Metal_enrichment} for explanation), which causes a change on the speed of enrichment (but with a 10 Myr delay). The serrated shape is caused by our simplification that all the stars formed in a 10 Myr star formation epoch have an identical age.}
        \label{fig:Z_over_X_time}
    \end{figure}

    Fig.~\ref{fig:Z_over_X_time} shows that the metal enrichment happens at an earlier time under the IGIMF theory relative to the invariant canonical IMF. It follows that the IGIMF theory results in earlier metal enrichment and higher final metallicity. The metal abundances of galaxies during the first few hundred Myr are difficult to constrain but the next generation James Webb Space Telescope may allow such observation of the high-redshift universe.
    
    Note that the stars only form at 10 Myr time steps. The temporary decrease of the mass- and luminosity-weighted stellar metallicity during the 10 Myr epoch is caused by the death of the most recently formed metal-rich massive stars.
    
    \subsubsection{Helium abundance}\label{sec:helium}
    
    \begin{figure}
        \centering
        \includegraphics[width=\hsize]{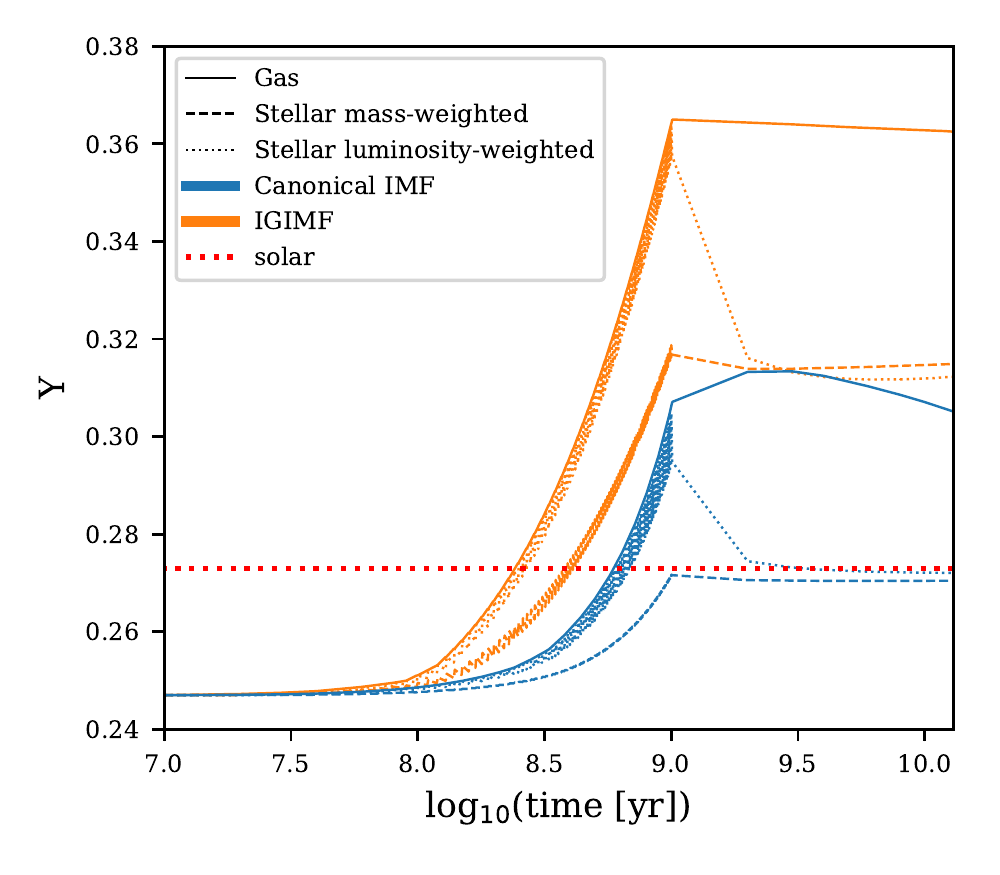}
        \caption{Same as Fig.~\ref{fig:Z_over_X_time} but for helium mass fraction, Y. The Solar value of 0.273 is adopted from \cite{2010ApJ...719..865S}.}
        \label{fig:Y_time}
    \end{figure}
    Fig.~\ref{fig:Y_time} shows the enrichment of helium under different IMF assumptions and the helium-enrich is more prominent under the IGIMF theory. This is a natural outcome of the high helium yield of the massive stars according to the adopted stellar yield table as is shown in Fig.~\ref{fig:steller_yields}.
    
    It is important to self-consistently consider the helium abundance in galaxy chemical evolution models, i.e., when the galactic Y/Z changes, the adopted stellar yield should also come from a stellar evolution simulation with that initial Y/Z, because the abundance of carbon, nitrogen, and oxygen is highly influenced by the hydrogen burning process.
    
    \subsubsection{Iron abundance}\label{sec:Iron abundance}
    
    \begin{figure}
        \centering
        \includegraphics[width=\hsize]{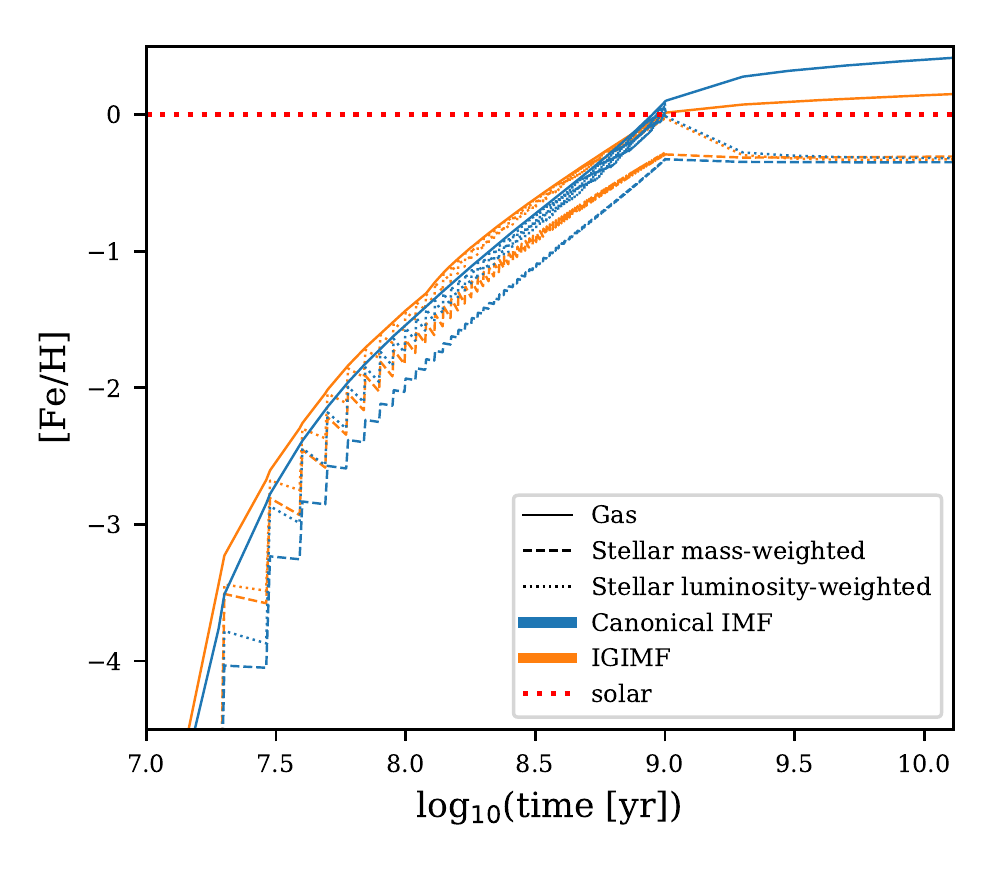}
        \caption{Same as Fig.~\ref{fig:Z_over_X_time} but for [Fe/H].}
        \label{fig:Fe_over_H_time}
    \end{figure}
    
    The iron evolution is shown in Fig.~\ref{fig:Fe_over_H_time}. We highlight three features of the result obtained using the IGIMF theory due to a larger fraction of iron produced by SNII relative to SNIa.
    
    \begin{itemize}
    \item The resulting [Fe/H] is lower than the [Z/X] (compare to the IGIMF line in Fig.~\ref{fig:Z_over_X_time}), consistent with the $\alpha$-enhanced characteristic of massive elliptical galaxies (see Section~\ref{sec:alpha-enhancement} below).
    
    \item The iron production is faster at earlier times thanks to the enhanced number of SNII but slower at later times (see Fig.~\ref{fig:SN_number_evolution} above). The earlier production of iron is in line with the suggestion from the abundance profiles of the ICM \citep{2017A&A...603A..80M}.
    
    \item The iron abundance difference at 13 Gyr between the gas phase and stellar phase is smaller compared to the models resting on the canonical IMF assumption, alleviating the difficulty in explaining the observed equality of stellar and gas-phase iron abundance (see \citealt{2011A&A...530A..98P}).
    \end{itemize}
    
    \subsubsection{Alpha element enhancement}\label{sec:alpha-enhancement}
    
    \begin{figure}
        \centering
        \includegraphics[width=\hsize]{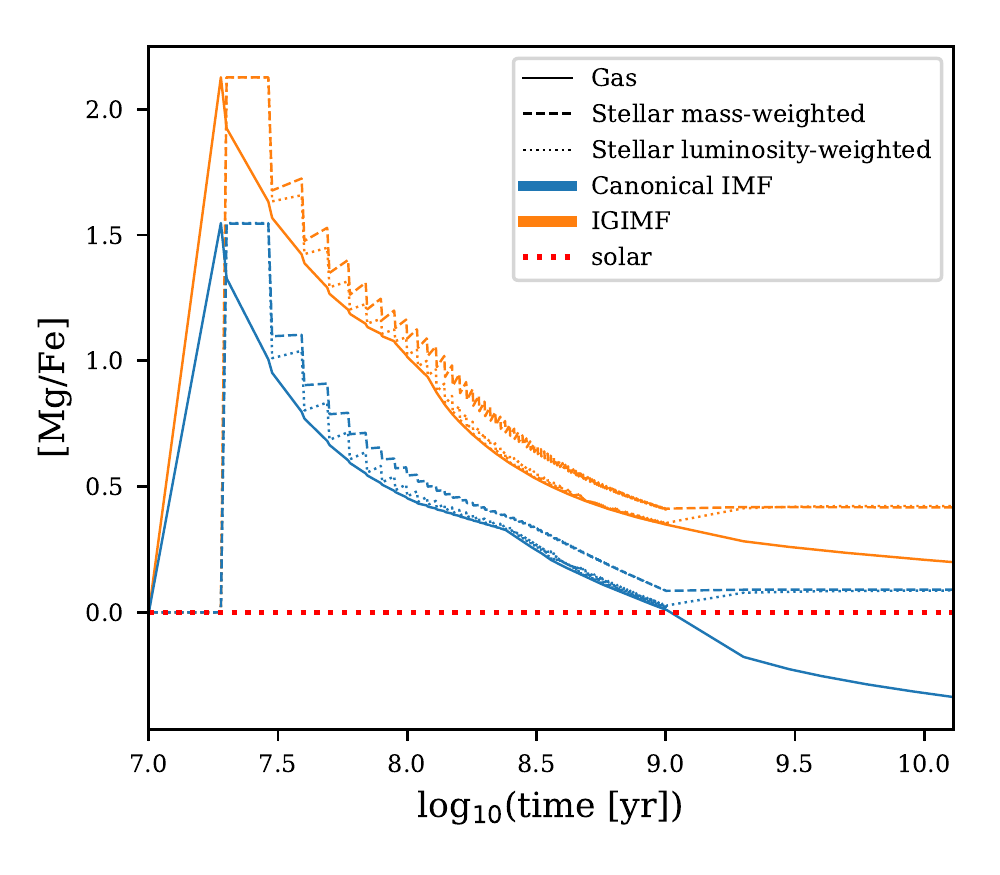}
        \caption{Same as Fig.~\ref{fig:Z_over_X_time} but for [Mg/Fe]. The discontinuous change of slope of the curves, e.g., the blue curves near $10^{8.4}$ yr, is explained in Fig.~\ref{fig:Z_over_X_time} with the horizontal "Yield Switch" lines.}
        \label{fig:Mg_over_Fe_time}
    \end{figure}
    Fig.~\ref{fig:Mg_over_Fe_time} shows that IGIMF-applied galaxy evolution results in a higher [Mg/Fe], in line with the results obtained by \citet{2009A&A...499..711R}. 
    
    Note that \citet{2009A&A...499..711R} applied an early version of the IGIMF theory where the IMF variation of individual star clusters was not applied. The modern IGIMF prescription used here with a possible top-heavy gwIMF further increases the resulting [Mg/Fe] value.
    
\section{Conclusions}\label{sec:conclusion}

    This contribution publishes an open source galaxy evolution code that is able to vary the gwIMF according to the IGIMF theory (or any other variable IMF theory) at each time step, yielding the mass composition, supernova rate, and chemical composition evolution of the galaxy. Following \citet{2017A&A...607A.126Y} and \citet{2018A&A...620A..39J},
    %added this as in the second paper we have also metallicity dependency, thus it is important for completeness.
    where the gwIMF for a single star formation epoch is derived according to the IGIMF theory, the newly developed code published here enables us to, for the first time, perform a comprehensive and detailed examination of the effect of applying a different IMF theory in the context of galaxy chemical evolution assuming an identical SFH.
    
    Our preliminary result of a high-mass starburst galaxy assuming different IMFs and the same SFH is shown in Section~\ref{sec:result}; while the result assuming the same gwIMF (the IGIMF theory) but different SFHs is shown in Appendix~\ref{sec:lognorm_results}.
    A systematically varying gwIMF leads to a strong modification of the element composition of a galaxy (naturally reproducing the observed higher metallicity and $\alpha$-enhancement of the massive elliptical galaxies); while the difference in SFH and recent star formation activity influences the interpreted property of a galaxy due to a difference in the luminosity-weighted observation. 
    
    The star formation environment of massive ellipticals in the early Universe is very different from the local Universe where we calibrate the IGIMF theory. There is no guarantee that such an extrapolation of the theory would work but it is certainly an inspiring step. We will use this open source code for more comprehensive and detailed studies for galaxies with different masses and SFHs and also develop a spatially resolved IGIMF formulation.
    
\begin{acknowledgements}
    ZY acknowledges financial support from the China Scholarship Council (CSC). TJ acknowledges support through an ESO studentship. We acknowledge the open source NuPyCEE project from where we adopt the digital stellar yield tables. This work also benefited from the International Space Science Institute (ISSI/ISSI-BJ) in Bern and Beijing, thanks to the funding of the team “Chemical abundances in the ISM: the litmus test of stellar IMF variations in galaxies across cosmic time” (Principal Investigator Donatella Romano and Zhi-Yu Zhang). We thank especially Donatella Romano for reading our manuscript and providing important comments. Last but not lest, we thank Daniel Thomas for his helpful comments that improved our paper.
\end{acknowledgements}

\bibliographystyle{aa} % style aa.bst
\bibliography{references}

\begin{appendix}

\section{Results for a log-normal SFH}\label{sec:lognorm_results}
    
    This section compares the simulation results applying the IGIMF theory but with the two different SFHs shown in Fig.~\ref{fig:SFH}. The log-normal SFH forms the same stellar mass but in a more extended timescale. The tail of the SFH with a low SFR leads to a stellar population being contributed every $\delta t = 10\,$Myr with a top-light gwIMF which does not exist in the box-shaped SFH model used in the main text. Thus a more realistic and extended SFH naturally leads to a time-varying gwIMF suggested qualitatively already by \citet{1996ApJS..106..307V,1997ApJS..111..203V,2013MNRAS.435.2274W,2015MNRAS.448L..82F,2016MNRAS.462.3854L}, with a top-heavy gwIMF in the beginning and a bottom-heavy (and top-light) gwIMF at later times.
    
\begin{figure}[ht]
        \centering
        \includegraphics[width=\hsize]{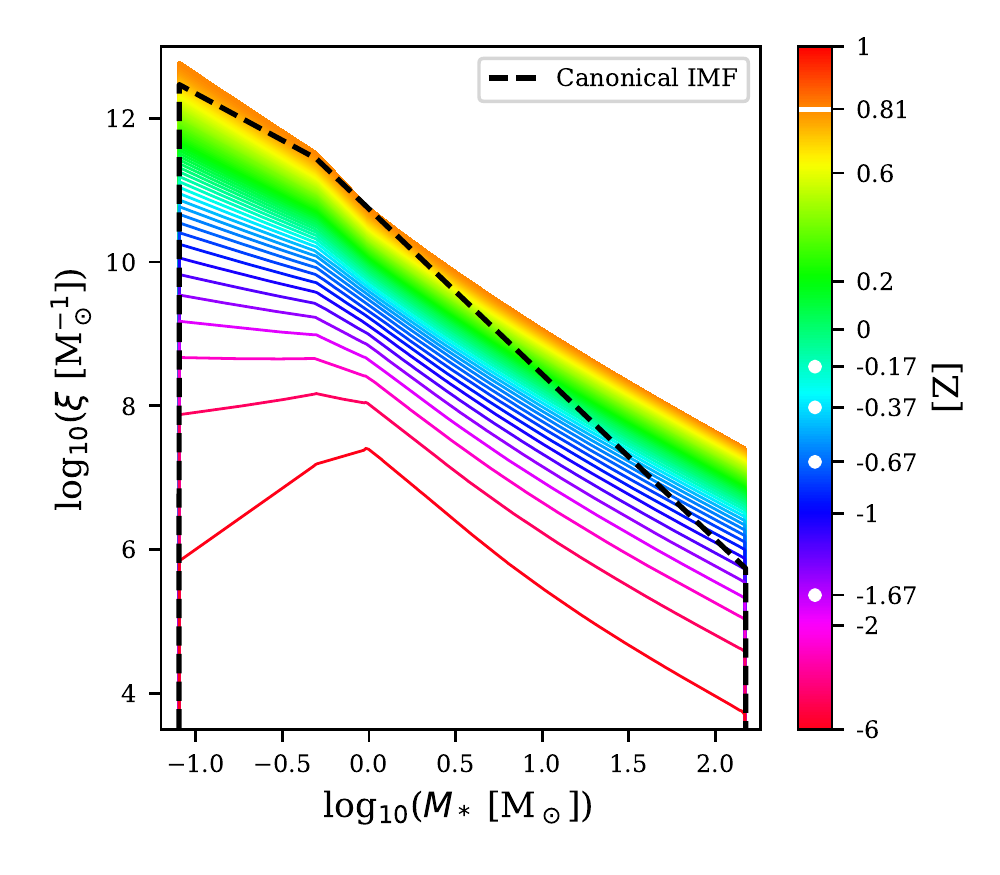}
        \caption{Same as Fig.~\ref{fig:1000_IMF_evolution} but for the log-normal SFH shown in Fig.~\ref{fig:SFH} that extends for 13 Gyr.}
        \label{fig:extended_1000_IMF_evolution}
    \end{figure}

    Fig.~\ref{fig:extended_1000_IMF_evolution} shows the resulting TIgwIMF for the log-normal SFH. The variation of the low-mass part of the TIgwIMF is smoother than the boxy SFH case (Fig.~\ref{fig:1000_IMF_evolution}) because the metal enrichment is relatively slower as is shown in Fig.~\ref{fig:extended_Z_over_X_time}.
    
\begin{figure}[ht]
        \centering
        \includegraphics[width=\hsize]{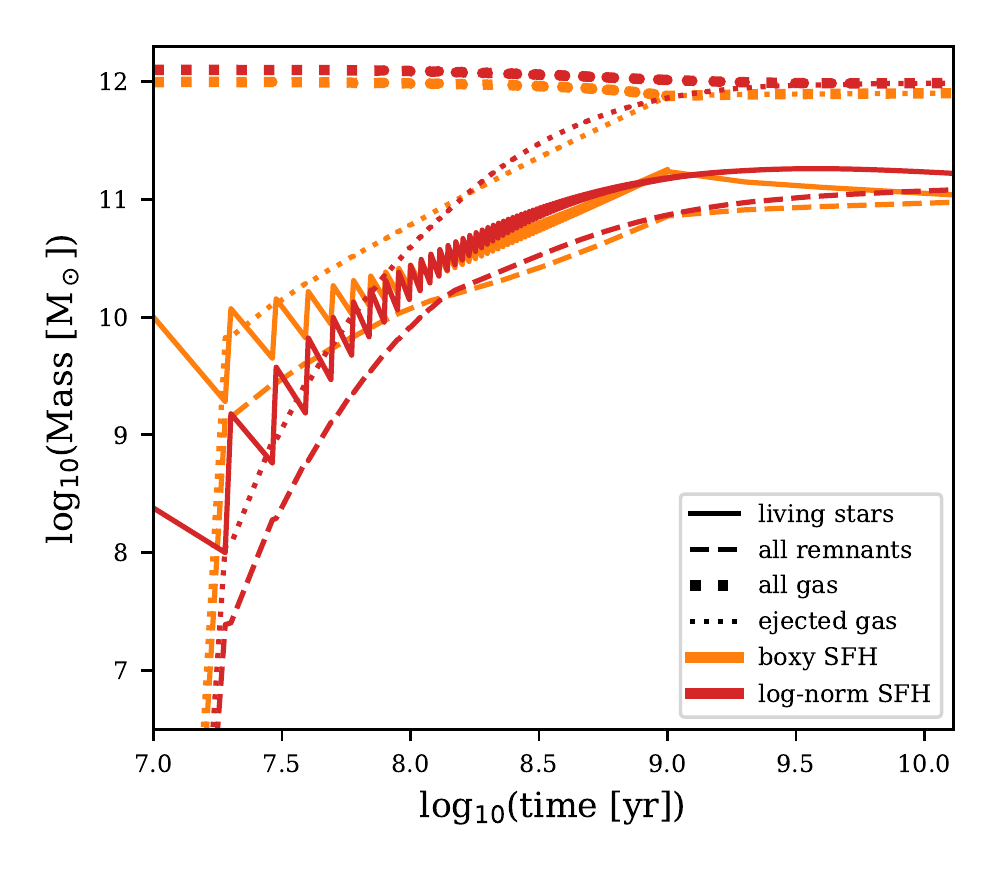}
        \caption{Same as Fig.~\ref{fig:mass_evolution} but comparing different SFHs instead of different IMF assumptions. The orange lines apply the boxy SFH and the red lines apply the log-normal SFH, both shown in Fig.~\ref{fig:SFH}. The same color code is applied from Fig.~\ref{fig:extended_mass_evolution} to \ref{fig:extended_Mg_over_Fe_time}. The serrated shape is caused by the dying stars with a lifetime shorter than 10 Myr (see Section~\ref{sec:mass_evolution}). Note that the orange lines are the same as the orange lines in Fig.~\ref{fig:mass_evolution}.}
        \label{fig:extended_mass_evolution}
    \end{figure}

\begin{figure}
        \centering
        \includegraphics[width=\hsize]{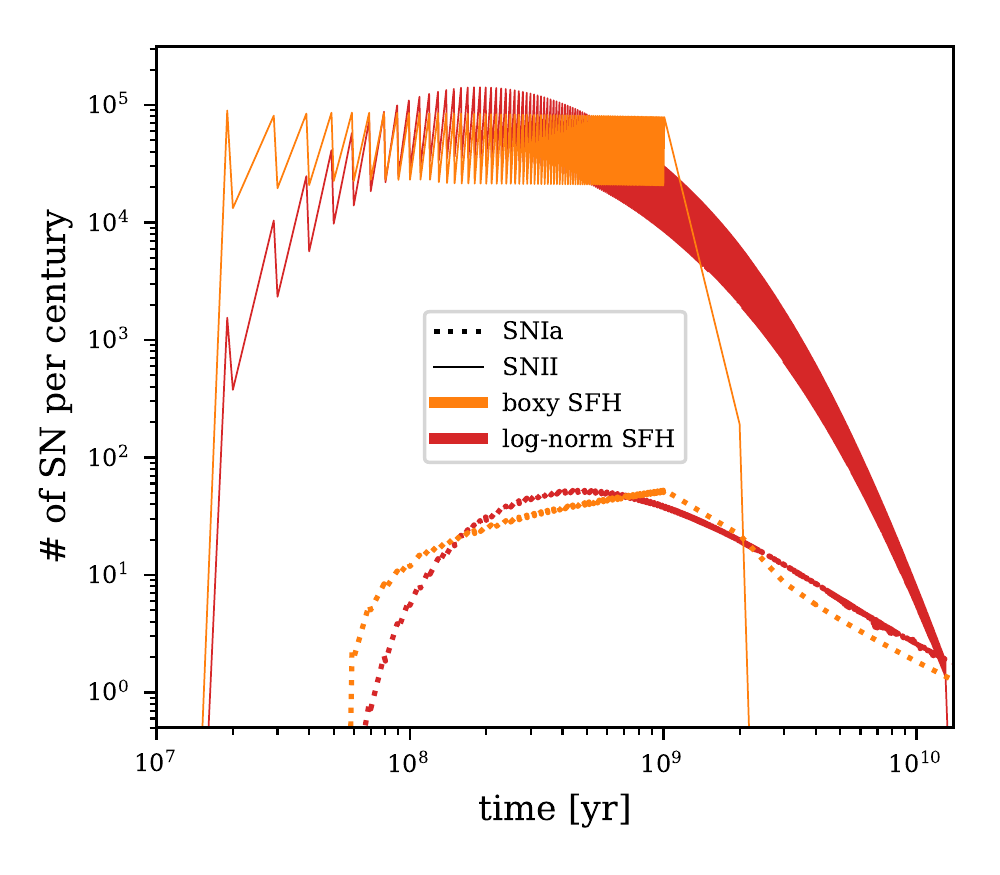}
        \caption{Same as Fig.~\ref{fig:extended_mass_evolution} and Fig.~\ref{fig:SN_number_evolution}. The orange lines are the same as the orange lines in Fig.~\ref{fig:SN_number_evolution}.}
        \label{fig:extended_SN_number_evolution}
    \end{figure}

\begin{figure}
        \centering
        \includegraphics[width=\hsize]{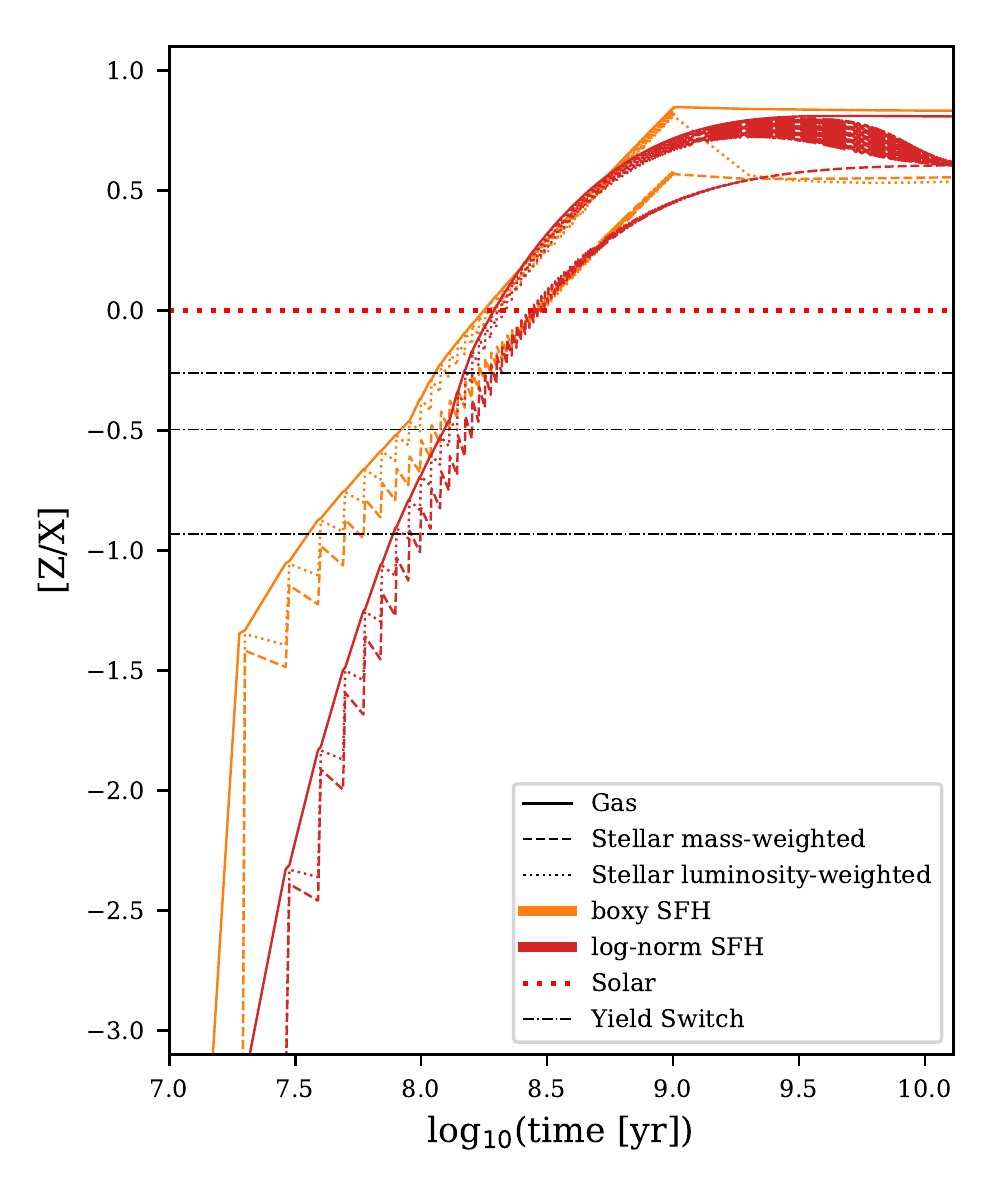}
        \caption{Same as Fig.~\ref{fig:extended_mass_evolution} and Fig.~\ref{fig:Z_over_X_time}. The orange lines are the same as the orange lines in Fig.~\ref{fig:Z_over_X_time}.}
        \label{fig:extended_Z_over_X_time}
    \end{figure}

\begin{figure}
        \centering
        \includegraphics[width=\hsize]{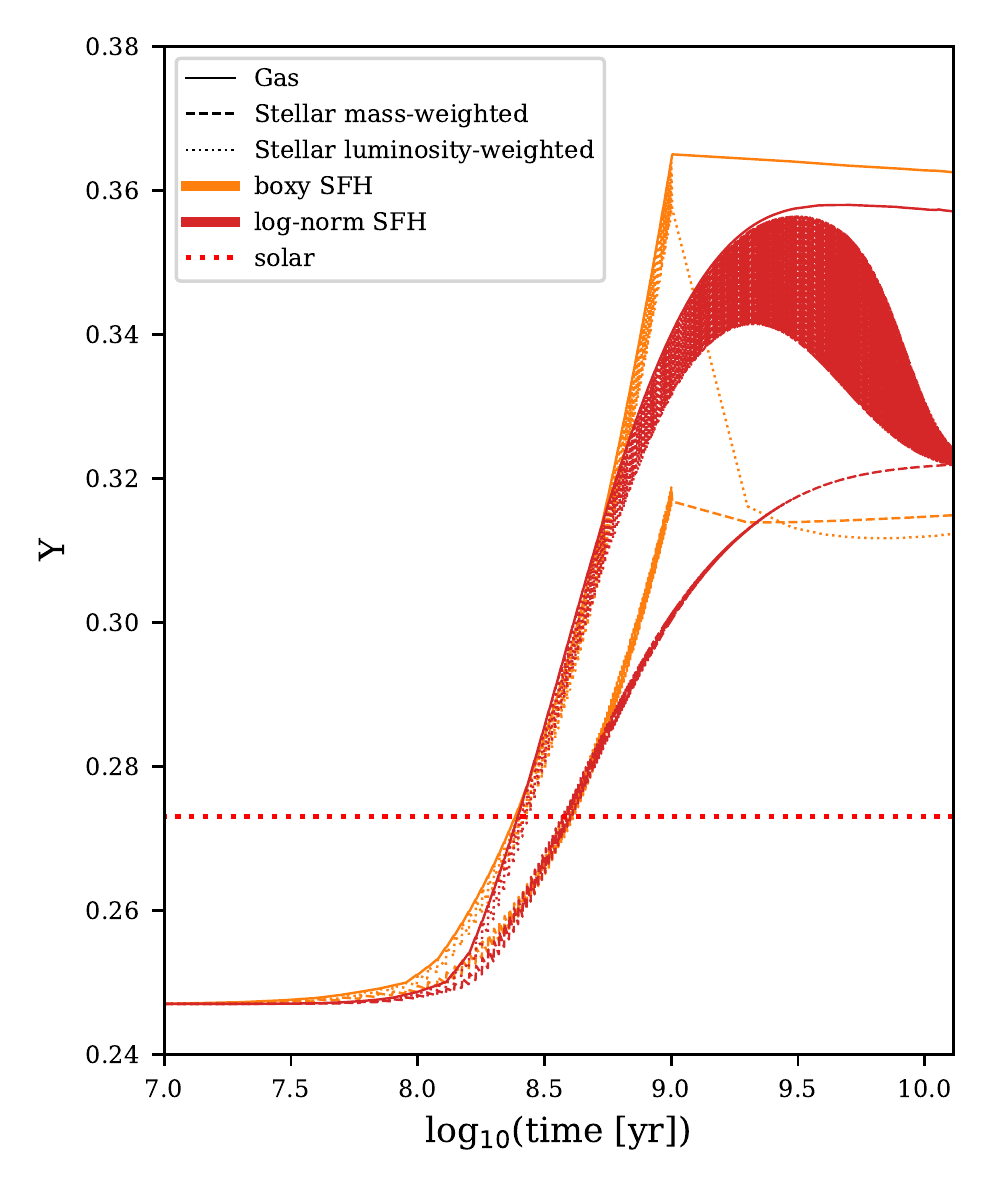}
        \caption{Same as Fig.~\ref{fig:extended_mass_evolution} and Fig.~\ref{fig:Y_time}. The orange lines are the same as the orange lines in Fig.~\ref{fig:Y_time}.}
        \label{fig:extended_Y_time}
    \end{figure}

\begin{figure}
        \centering
        \includegraphics[width=\hsize]{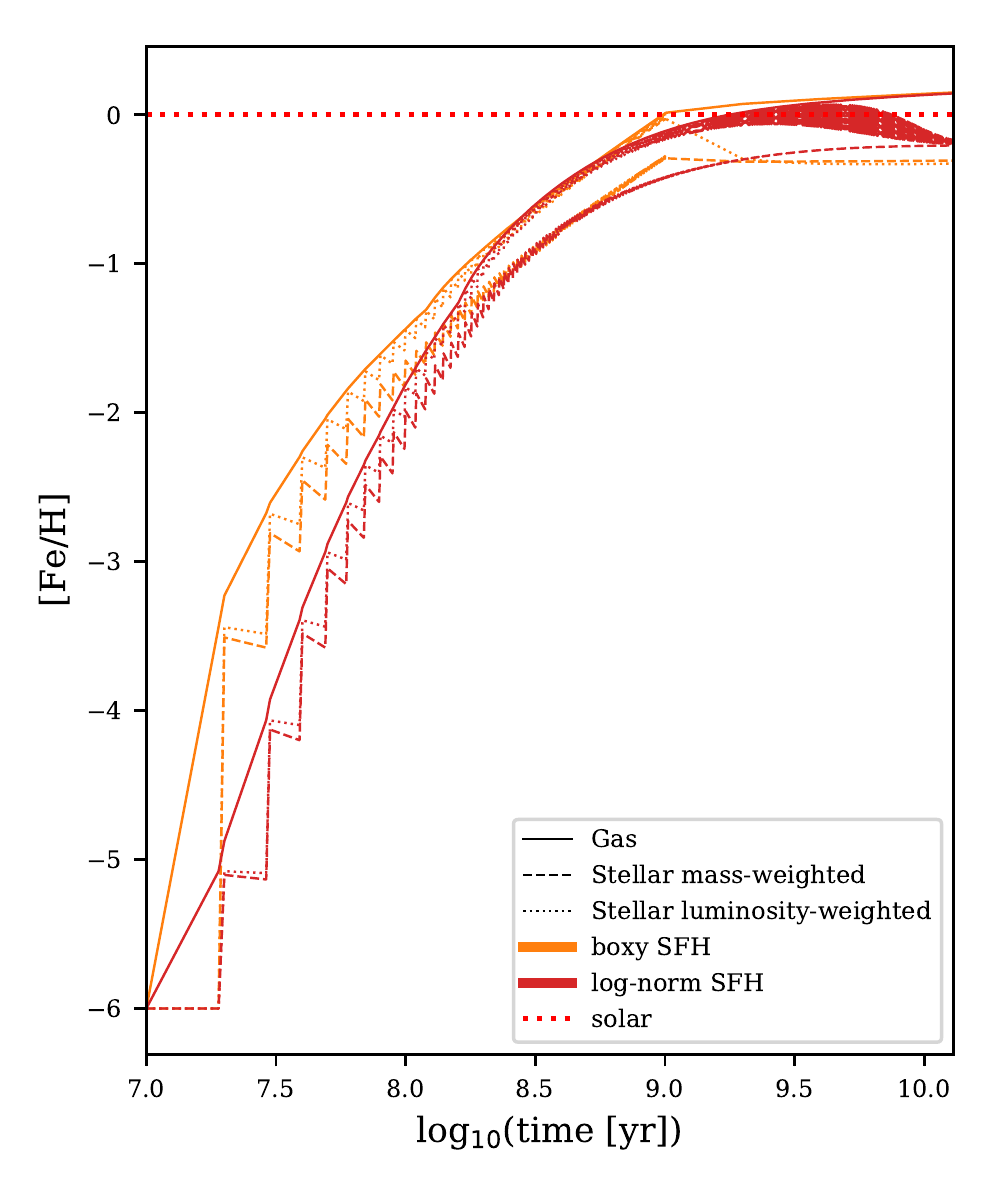}
        \caption{Same as Fig.~\ref{fig:extended_mass_evolution} and Fig.~\ref{fig:Fe_over_H_time}. The orange lines are the same as the orange lines in Fig.~\ref{fig:Fe_over_H_time}.}
        \label{fig:extended_Fe_over_H_time}
    \end{figure}

\begin{figure}
        \centering
        \includegraphics[width=\hsize]{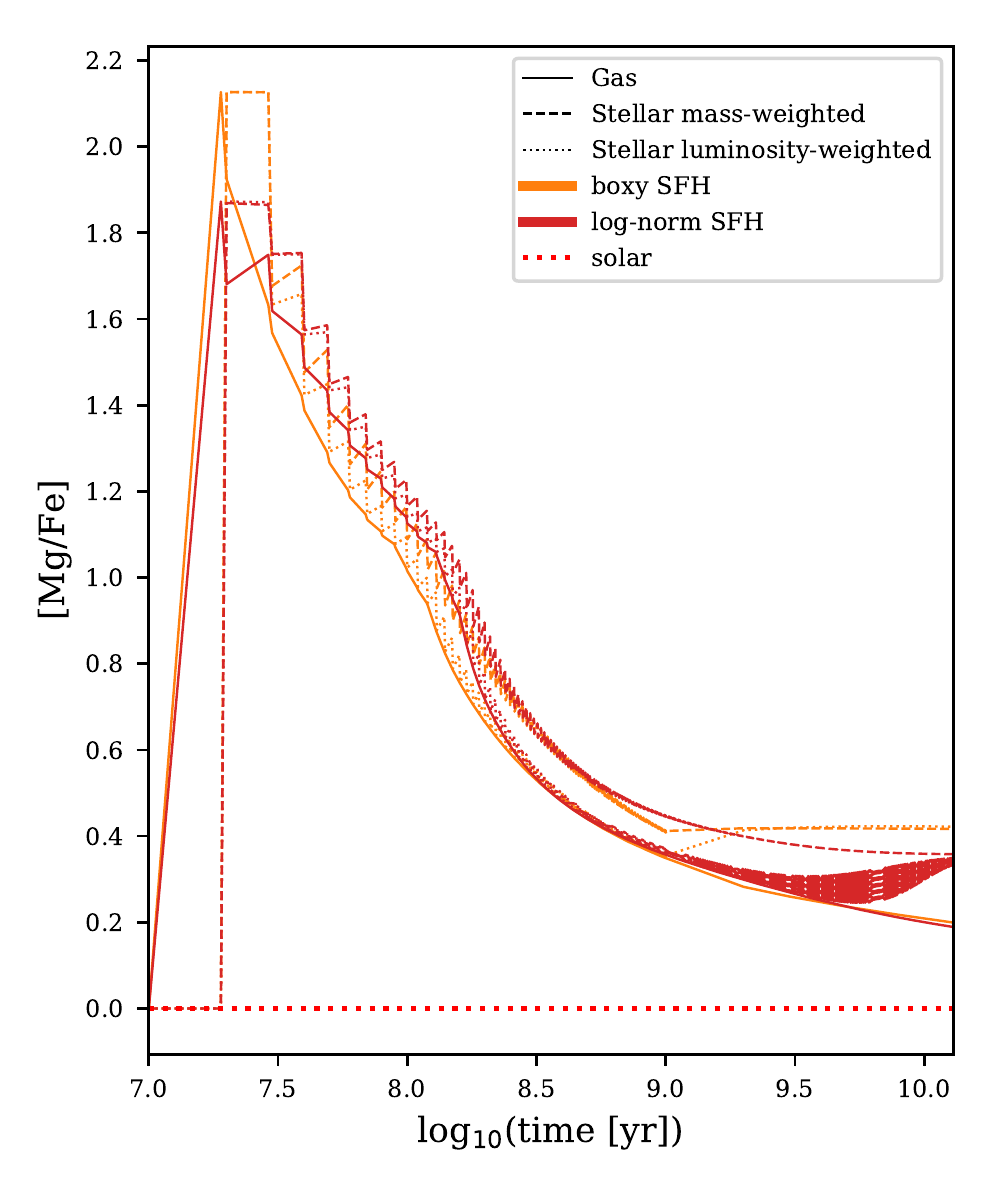}
        \caption{Same as Fig.~\ref{fig:extended_mass_evolution} and Fig.~\ref{fig:Mg_over_Fe_time}. The orange lines are the same as the orange lines in Fig.~\ref{fig:Mg_over_Fe_time}.}
        \label{fig:extended_Mg_over_Fe_time}
    \end{figure}

    Fig.~\ref{fig:extended_mass_evolution} to \ref{fig:extended_Mg_over_Fe_time} apply the same color coding of Fig.~\ref{fig:SFH}. The orange lines here labeled with "boxy SFH" are the same as the orange lines in Fig.~\ref{fig:mass_evolution} to \ref{fig:Mg_over_Fe_time} labeled with "IGIMF". The luminosity-weighted stellar element abundance becomes closer to the gas-phase value as the log-normal SFH has a more recent star formation activity. The luminosity-weighted result depends on the SFH and fluctuates for different galaxies that have different SFHs.
    
\section{Luminosity-weighted element ratio}\label{sec:Luminosity-weighted element ratio}

    The luminosity-weighted element ratio, e.g., [Fe/H]$_{\rm lw}$ is not weight with the [Fe/H] of each star but with the element mass, $M_{i, \rm mw}$, of the stars, where $i$ is Fe or H and
    \begin{equation}
    M_{i, \rm mw}(t_n) = \frac{\sum\limits_{t=t_1}^{t_n} \left[ \int_{m=m_{\rm min}}^{m_{\rm max}(t_n-t)} \xi(t,M_{*,\rm initial})L(M_{*,\rm initial})dm \cdot M_{i}(t) \right]}{\sum\limits_{t=t_1}^{t_n} \left[ \int_{m=m_{\rm min}}^{m_{\rm max}(t_n-t)} \xi(t,M_{*,\rm initial})L(M_{*,\rm initial})dm \right]},
    \end{equation}
    then calculate the abundance ratio with
    \begin{equation}
    \begin{split}
    \mathrm{[Fe/H]}_{\rm mw}(t_n) =& \mathrm{log}_{10}[(M_{\rm Fe, mw}(t_n)/W_{\rm Fe}) / (M_{\rm H, mw}(t_n)/W_{\rm H})] \\
           &- \mathrm{log}_{10}(A_{\rm Fe} / A_{\rm H}),
    \end{split}
    \end{equation}
    where $A_{\rm i}=\mathrm{log}_{10}(N_{i}/N_{\rm H})+12$ is the solar photospheric abundance of element $i$ adopted from \citet[their Table 2]{1989GeCoA..53..197A}.
    
\end{appendix}

\end{document}